\documentclass[11pt]{article}
\RequirePackage[OT1]{fontenc}
\RequirePackage{amsthm,amsmath,natbib}
\RequirePackage[colorlinks,citecolor=blue,urlcolor=blue]{hyperref}
\usepackage{amssymb,amsmath,latexsym}
\usepackage[T1]{fontenc}
\usepackage{enumerate,url}
\usepackage{adjustbox}
\usepackage{graphicx}
\usepackage{enumitem}
\usepackage{ulem}

\newcommand{\bbeta}{\boldsymbol{\beta}}

\newcommand{\bepsilon}{\boldsymbol{\epsilon}}
\newcommand{\beeta}{\boldsymbol{\eta}}

\newcommand{\blambda}{\boldsymbol{\lambda}}
\newcommand{\bLambda}{\boldsymbol{\Lambda}}
\newcommand{\bvartheta}{\boldsymbol{\vartheta}}
\newcommand{\btheta}{\boldsymbol{\theta}}

\newcommand{\bPsi}{\boldsymbol{\Psi}}

\newcommand{\bzero}{\boldsymbol{0}}

\newcommand{\bdelta}{\boldsymbol{\delta}}

\newcommand{\bSigma}{\boldsymbol{\Sigma}}

\newcommand{\bomega}{\boldsymbol{\omega}}
\newcommand{\bOmega}{\boldsymbol{\Omega}}

\newcommand{\ba}{\mathbf{a}}

\newcommand{\be}{\boldsymbol{e}}
\newcommand{\bff}{\boldsymbol{f}}

\newcommand{\bu}{\boldsymbol{u}}
\newcommand{\bp}{\boldsymbol{p}}
\newcommand{\bA}{\boldsymbol{A}}
\newcommand{\bB}{\boldsymbol{B}}

\newcommand{\bc}{\boldsymbol{c}}
\newcommand{\bD}{\boldsymbol{D}}
\newcommand{\bF}{\boldsymbol{F}}

\newcommand{\bh}{\boldsymbol{h}}

\newcommand{\bI}{\boldsymbol{I}}
\newcommand{\bJ}{\boldsymbol{J}}

\newcommand{\bm}{\boldsymbol{m}}

\newcommand{\bR}{\boldsymbol{R}}
\newcommand{\bQ}{\boldsymbol{Q}}
\newcommand{\bS}{\boldsymbol{S}}

\newcommand{\bx}{\boldsymbol{x}}
\newcommand{\bX}{\boldsymbol{X}}

\newcommand{\by}{\boldsymbol{y}}

\newtheorem{theo}{Theorem}
\newtheorem{lem}{Lemma}

\newtheorem{rem}{Remark}
\newtheorem{exemple}{Example}
\numberwithin{equation}{section}
\numberwithin{theo}{section}
\numberwithin{lem}{section}
\numberwithin{cor}{section}
\numberwithin{exemple}{section}
\numberwithin{rem}{section}
\numberwithin{prop}{section}
\renewcommand{\baselinestretch}{1.5}\small\normalsize
\def\zak{\null\hfill{$\Box$}\par\vspace*{0.2cm}}

\def\1g{1\hskip -3pt \mbox{l}}

\title{Virtual Historical Simulation for estimating  the conditional VaR of large portfolios\\
}

\author{
{\sc Christian Francq\footnote{CREST and University of Lille, BP 60149, 59653 Villeneuve d'Ascq cedex, France. E-Mail:
christian.francq@univ-lille3.fr} and Jean-Michel
Zakoïan\footnote{Corresponding author: Jean-Michel Zakoïan,
University of Lille  and CREST, 5 Avenue Henri Le Chatelier, 91120 Palaiseau, France. E-mail: zakoian@ensae.fr.
 } }}

\begin{document}
\renewcommand{\baselinestretch}{1.5}
%
%
\date{}
\maketitle

\vspace{-1cm}
\begin{quote}
\begin{center}
\textbf{Abstract}
\end{center}
{\small
\hspace{1em}
In order to estimate the conditional risk of a portfolio's return, two strategies can be advocated. A multivariate strategy requires estimating a dynamic model for the vector of risk factors,
which is often challenging, when at all possible, for large portfolios.
A univariate approach based on a dynamic model for the portfolio's return seems more attractive.
However, when the combination of the individual returns is time varying, the portfolio's return series is typically non stationary which may
         invalidate statistical inference. An alternative approach consists in reconstituting a "virtual portfolio", whose returns are built using the current composition of the portfolio and for which a stationary dynamic model can be estimated.
         This paper establishes the asymptotic properties of this method, that we call Virtual Historical Simulation. Numerical illustrations on simulated and real data are provided.
}
\end{quote}



\renewcommand{\baselinestretch}{1.5}\small\normalsize

\newpage
\section{Introduction}

The quantitative standards laid down under Basel Accord II and III
allow banks to develop internal models for setting aside capital.
Methods that incorporate time dependence to quantify market risks are able to use
knowledge of the conditional distribution.
In particular, the conditional Value-at-Risk (VaR)
of financial returns, with a given risk level $\alpha$ (typically, $\alpha=1\%$ or $5\%$)  is nothing else,
from a statistical point of view, than the negated
$\alpha$-quantile of the conditional distribution of the portfolio returns.
Estimating conditional quantiles, or more generally conditional risk measures, of a time series of financial returns is thus crucial for risk management.

It is also essential, for risk management purposes,  to be able to evaluate the accuracy of such estimators of conditional risks.
Uncertainty  implied by statistical procedures in the implementation of risk measures
may lead to false security in financial markets (see e.g. Farkas, Fringuellotti and Tunaru (2016) and the references therein).
Estimation risk thus needs to be accounted for, in addition to market
risk.
However, 
evaluating the estimation risk
for the conditional Value-at-Risk (VaR)
is generally challenging for two main reasons.
Firstly, because the stochastic nature of the conditional VaR does not allow in general
to reduce the problem to the estimation of a parameter. Making inference on a stochastic process is obviously more intricate than on a parameter.
Secondly, quantiles being obtained as the solutions of optimization problems based on non-smooth functions, establishing asymptotic properties of conditional VaR estimators may become a difficult task.

Increasing attention has been directed in the recent econometric literature to the inference of risk measures in dynamic risk models.
Francq and Zakoïan (2015) derived asymptotic confidence intervals (CI) for the conditional VaR of a series of financial returns driven by a parametric dynamic model.
Robust backtesting procedures were developed by Escanciano and Olmo (2010, 2011),
and Gouriéroux and Zakoïan (2013) studied the effect of estimation
on the coverage probabilities. Several articles proposed resampling methods: among others, Christoffersen and Gonçalves (2005) and Spierdijk (2016) considered using bootstrap procedures
for constructing CIs for VaR;
Hurlin, Laurent, Quaedvlieg and Smeekes (2017) proposed bootstrap-based comparison tests of two conditional risk measures.
Beutner, Heinemann and  Smeekes (2019) established the validity of a fixed-design residual bootstrap method for the two-step conditional VaR estimator of Francq and Zakoïan (2015).
See Nieto and Ruiz (2016) for an extensive survey of the methods for constructing and evaluating VaR forecasts that have been proposed in the literature.

Most existing studies on risk measure inference focus on  the risk of a single financial asset.
The aim of the present article is to estimate conditional VaR's for {\it portfolios} of financial assets. From a statistical point of view, the extension is far from trivial.
First, because evaluating the quantile of a linear combination of variables may require knowledge of the complete joint distribution
of such variables. When the object of interest is a {\it conditional} quantile, this approach requires specifying a dynamic model for the vector of returns of the assets
involved in the portfolio.
Second, portfolios compositions are generally time-varying, in particular if the agents adopt a mean-variance approach which, in a dynamic framework, requires specifying the first two conditional
moments. This typically entails non-stationarity of the portfolio's return time series, as we shall see in more detail.

A natural approach for obtaining the conditional VaR of a portfolio relies on specifying a multivariate GARCH model
for the vector of underlying asset returns.
Rombouts and Verbeek (2009) proposed a semi-parametric approach relying on estimating the conditional density of the innovations vector and evaluating numerically
the conditional VaR of a portfolio.
The asymptotic properties of similar multivariate  methods--with or without the assumption of sphericity of the innovations vector--were investigated by Francq and Zakoïan (2018).
As noted by Rombouts and Verbeek the advantage of  multivariate approaches is to "take into account the dynamic interrelationships between the portfolio components, while the model
underlying the VaR calculations is independent of the portfolio composition". On the other hand, for large portfolios multivariate approaches often become untractable due to the well-known dimensionality curse.

In this paper, we consider {\it univariate} procedures aiming at handling portfolios constructed with a large number of assets.
We first
consider
a "naive" approach in which a standard volatility model is directly fitted to the
portfolio returns. Despite its empirical relevance, we will see that
the naive approach is not amenable to asymptotic statistical inference
(due to the inherent non stationarity of the observed portfolio's time series).
We
study the asymptotic properties of an alternative 
procedure relying on a "virtual portolio" constructed with the {\it current composition} of the portfolio, on which a univariate model is fitted.
This procedure--which we call Virtual Historical Simulation (VHS)--is amenable to asymptotic statistical inference. From a numerical point of view, it allows to avoid difficulties caused by the dimensionality curse
in estimation of multivariate volatility models for vectors of asset returns.

The VHS method is related to other approaches introduced in Finance.
The Basel Committee and European Union directives (UCITS) recommend that banks backtest their VaR measures against both "clean" and "dirty" P\&L's of their trading portfolios (see Holton, 2014). Dirty P\&L's are the actual P\&L's reported at the end of the time horizon. They can be impacted by changes in the composition of the portfolio that occur during the VaR horizon. Since such position changes may have exogenous causes that can not be anticipated,  it is relevant to backtest the VaR with the so-called clean P\&L, which is the hypothetical  P\&L that would occur if the composition of the portfolio remained unchanged and if market moves were the only source of P\&L change (see Pérignon, Deng and Wang, 2008). Clean P\&L thus excludes  P\&L arising from intra-day trading, new trades, changes in reserves, fees and commissions.  Clean P\&L's are often used in the backtesting process, but they can also be used for VaR estimation. The VHS method exploits the idea of cleaning the P\&L's for computing the VaR. At each past date $t$, one can compute a virtual return--the opposite of a clean (or hypothetical) P\&L--that would occur if day $t$ positions 
were exactly those of the current date. Even if each bank uses its own internal VaR model, most financial institutions compute VaR through filtered or simple historical simulations 
on plain or hypothetical (virtual) returns (see Laurent and Omidi Firouzi, 2017). This is the aim of the present paper to study the asymptotic properties of such VaR evaluation methods.

The paper is organized as follows.
Section \ref{secsetup} defines 
the
conditional VaR of a portfolio whose composition at the current date may depend on the historical prices, and presents
the naive and VHS estimation methods.
In Section \ref{secVal} we derive the
asymptotic properties of the VHS procedure based on the Gaussian
Quasi-Maximum Likelihood (QML) criterion,  under general assumptions on the volatility model.
Section \ref{secillus} presents some numerical illustrations based on Monte Carlo experiments and real financial data.
Proofs are collected in the Appendix.

\section{Estimating the conditional VaR}\label{secsetup}

\subsection{Conditional VaR of a dynamic portfolio}

Let $\bp_t=(p_{1t}, \ldots, p_{mt})'$ denote the vector of prices of
$m$ assets at time $t$. Let $\by_t=(y_{1t}, \ldots,
y_{mt})'$ denote the corresponding vector of log-returns,
with $y_{it}=\log(p_{it}/p_{i,t-1})$ for $i=1, \ldots, m$.

 Let $V_t$ denote the value at time $t$ of a portfolio 
 composed of $\mu_{i,t-1}$
units of asset $i$, for $i=1, \ldots, m$:
\begin{equation}\label{V}
V_0=\sum_{i=1}^m \mu_{i}p_{i0},\quad
V_t=\sum_{i=1}^m \mu_{i,t-1}p_{it}, \quad \mbox{for } t\geq 1
\end{equation}
where the $\mu_{i,t-1}$ are measurable functions of the prices up to time $t-1$, and
the $\mu_{i}$ are constants.
The return of the portfolio over the period $[t-1,t]$ is, for  $t\geq 1$, assuming that
$V_{t-1}\ne 0,$
$$\frac{V_{t}}{V_{t-1}}-1=\sum_{i=1}^m a_{i,t-1}e^{y_{it}}-1\approx
\sum_{i=1}^m a_{i,t-1}y_{it}+a_{0,t-1}$$
where
$$a_{i,t-1}=\frac{\mu_{i,t-1}p_{i,t-1}}{\sum_{j=1}^m \mu_{j,t-2}p_{j,t-1}}, \quad i=1, \ldots,
m \quad \mbox{and}\quad a_{0,t-1}=-1+\sum_{i=1}^ma_{i,t-1}.$$
We assume that,
at date $t$, the investor may rebalance his portfolio under a "self-financing"
constraint.
\begin{itemize}
\item[\hspace*{0.2em} {\bf SF:}]
\hspace*{1em} The portfolio is rebalanced in such a way that
$\sum_{i=1}^m \mu_{i,t-1}p_{it}=\sum_{i=1}^m \mu_{i,t}p_{it}.$
\end{itemize}
In other words, the value at time $t$ of the portfolio bought at time $t-1$ equals the value at time
$t$ of the portfolio bought at time $t$. An obvious consequence of the self-financing assumption {\bf SF}, is that
the change of value of the portfolio between $t-1$  and $t$ is only due to the change of value of the underlying assets:
$$V_{t}-V_{t-1}=\sum_{i=1}^m \mu_{i,t-1}(p_{i,t}-p_{i,t-1}).$$
Another consequence is that the weights $a_{i,t-1}$ sum up to 1, that is $a_{0,t-1}=0$. Thus, under {\bf SF} we have
$\frac{V_{t}}{V_{t-1}}-1\approx
r_{t}$, where
\begin{equation}\label{portfo}
r_{t}=\sum_{i=1}^m a_{i,t-1}y_{it}=\mathbf{a}'_{t-1}\by_t,\qquad a_{i,t-1}=\frac{\mu_{i,t-1}p_{i,t-1}}{\sum_{j=1}^m \mu_{j,t-1}p_{j,t-1}},\end{equation}
for $i=1, \ldots, m,$
and $\mathbf{a}_{t-1}=(a_{1,t-1}, \ldots, a_{m,t-1})'$.
A portfolio is usually called {\it crystallized}  when the number of units of each asset is time independent, that is
$\mu_{i,t-1}=\mu_i$ for each $i=1, \ldots, m$ and
for all $t$. We will call {\it static} a
portfolio with fixed proportion in value of each return, that is
when $a_{i,t-1}=a_i$
for each $i=1, \ldots, m$ and
 for all $t$.

The {\it
conditional} VaR of the portfolio's return process $(r_t)$ at risk level
$\alpha\in (0,1)$, denoted $\mbox{VaR}_{t-1}^{(\alpha)}(r_t)$, is characterized by
\begin{equation}\label{truevar}
\mbox{VaR}_{t-1}^{(\alpha)}(r_t)=\inf\{x \;:\; P_{t-1}(-r_t\le x)\ge 1-\alpha\}
\end{equation}
where $P_{t-1}$ denotes the historical distribution conditional on the  information $I_{t-1}$ available at time $t-1$.
The specification of $I_{t-1}$ will depend on the approach used. Multivariate approaches use full information, that is all past prices of all assets.
In the next approach we describe a univariate approach which only uses the past returns of the portfolio.

\subsection{
The naive approach}
\label{secInval}
A natural approach for evaluating the conditional VaR in (\ref{truevar})
when $I_{t-1}=\sigma(r_{s}, s<t)$ is to estimate a univariate GARCH model, or any time series model,
on the series of portfolio returns. We will see that this approach, which can be called "naive", may be misleading due to the fact
that the return's portfolio is a
time-varying combination of the individual returns.

For simplicity, we consider a crystallized
portfolio,
with weight $\mu_i$ and initial  price $p_{i0}$ for the asset $i\in\{1, \ldots, m\}$.
The composition $\ba_{t-1}$ of such a portfolio is non stationary in general.
Indeed, we have
$$\log \left(\frac{a_{i,t}}{a_{j,t}}\right)=\log\left(\frac{\mu_ip_{i,0}}{\mu_jp_{j,0}}\right)+\sum_{k=1}^t\Delta_{i,j,k}, \qquad \Delta_{i,j,k}=y_{i,k}-y_{j,k},$$
and $(\sum_{k=1}^t\Delta_{i,j,k})_{t\geq 1}$ is a non stationary integrated process of order 1 under general assumptions.\footnote{By the Chung-Fuks theorem, this is the case when $\by_t$ is iid with zero mean and a  non-singular covariance matrix $\bSigma$. The non stationarity of the process also holds, for instance, if the sequence $(\Delta_{i,j,k})_k$ is mixing and nondegenerated.}
More precisely, a consequence of the following lemma is that, with probability tending to one,  the composition $\ba_{t-1}$
of the portfolio converges to the set of the vectors $\be_i$ of the canonical basis (corresponding to single-asset portfolios): $P(\ba_{t-1}\in\{\be_1,\dots,\be_m\})\to 1$ as $t\to\infty$.
\begin{lem}
 \label{paire}
Consider a process $(\Delta_k)_{k\geq 1}$. Assume that there exist real sequences $a_n>0$ and $b_n$, both tending to zero, such that
\begin{equation}
\label{TCLgene}
Z_n:=a_n\sum_{k=1}^n\Delta_k+b_n\stackrel{{\cal L}}{\to} Z\mbox{ as }n\to\infty,
\end{equation}
for some random variable $Z$, whose cdf is continuous at 0 and such that $p=P(Z>0)\in (0,1)$.
 Then, for any $c>0$, we have $P(\sum_{k=1}^n\Delta_k>c)\to p$ and $P(\sum_{k=1}^n\Delta_k<-c)\to 1-p$ as $n\to\infty$.
\end{lem}
Note that a generalized central limit theorem of the form \eqref{TCLgene} holds for any iid sequence $(\Delta_k)$ whenever the distribution of $\Delta_k$ belongs to the domain of attraction of $Z$, which then follows a stable distribution.
If the assumptions of Lemma~\ref{paire} hold with $\Delta_k=\Delta_{i,j,k}$ for any pair $(i,j)$, with $i\neq j$, then all the ratios $a_{i,t}/a_{j,t}$ are arbitrarily close to either 1 or 0 with probability tending to 1 as $t\to\infty$. In that case, the composition $\ba_{t-1}$ tends to be totally undiversified, but is not always close to the same single-asset composition $\be_i$. If the dynamics of the individual returns $y_{it}$ are not identical, the dynamics of the return $r_t$ will be time-varying, and the naive method based on a fixed stationary GARCH model is likely to produce poor results.

Simulation experiments reported in Section \ref{secillus} confirm that for crystallized portfolios, the naive approach may behave badly due to the non stationarity of the univariate returns $r_t$.
Of course, for static portfolios the non stationarity issue vanishes, but such
portfolios may be considered as artificial\footnote{as it would require rebalancing every day in order to maintain a fixed composition in percentages.
In general, portfolios are rebalanced to minimize risk, leading to a non-static composition.}. The next section studies a remedy to the non stationarity issue, while keeping the univariate framework.

\subsection{The VHS approach}
 An alternative to the naive approach consists in reconstituting a "virtual portfolio", whose returns are built using the current composition of the portfolio.
Given the portfolio composition $\ba_{t_0-1}=\bx$, say, at  time $t_0$, 
we construct a process of virtual 
returns
$$r_{t}^*(\bx)=\bx'\by_t, \qquad t\in \mathbb{Z}$$
and we consider the information set $I_{t_0-1}=\sigma(r_{s}^*(\bx), s<t_0)$.
Note that, in general, $r_{t}^*(\bx)\ne r_t$ because the composition of the (non virtual) portfolio is time varying ($\ba_{t-1}\ne \bx$, in general, for $t\ne t_0$).
Given the stationarity of $(\by_t)$, it is clear that the series of virtual  returns $\{r_t^*(\bx)\}$ is also stationary, with
conditional moments
$$E_{t-1}[r_t^*(\bx)]
=:\mu_t(\bx), \qquad
\mbox{var}_{t-1}[r_t^*(\bx)]
=:\sigma_t^2(\bx),$$
where $E_{t-1}(X)=E(X\mid r_{s}^*(\bx), s<t)$ for any variable $X$, and the variance is defined accordingly.
Thus, $r_{t}^*(\bx)$ follows a model of the form
\begin{equation}\label{modVHS}
r_{t}^*(\bx)=\mu_t(\bx)+\sigma_t(\bx)u_t,\quad
\mbox{where $E_{t-1}(u_t)=0$ and $\mbox{var}_{t-1}(u_t)=1$.}
\end{equation}
Noting that $r_{t_0}=r^*_{t_0}(\ba_{t_0-1})$, the conditional VaR at time $t_0$ thus satisfies
\begin{equation}\label{complicatedVaR}
\mbox{VaR}_{t_0-1}^{*(\alpha)}(r_{t_0})=-\mu_{t_0}(\ba_{t_0-1})+\sigma_{t_0}(\ba_{t_0-1})\mbox{VaR}_{t_0-1}^{*(\alpha)}(u_{t_0})
\end{equation}
where $\mbox{VaR}_{t-1}^{*(\alpha)}(X)$ is the VaR of $X$ at level $\alpha$ conditional on $I_{t-1}.$

Note that the martingale difference $(u_t)$ may not be iid, as the following example illustrates. 
\begin{exemple}\label{noniid}
{\rm Consider the bivariate ARCH(1) process, defined as the stationary non anticipative solution of the model
$$\by_t=\left(\begin{array}{c}y_{1t}\\y_{2t}\end{array}\right)=
\bSigma_t\beeta_t, \quad
\bSigma_t=\left(\begin{array}{cc}\sigma_{1t}^2:= \omega_1+\alpha_{11}y_{1,t-1}^2+\alpha_{12}y_{2,t-1}^2&0\\0&\sigma_{2t}^2:= \omega_2+\alpha_{21}y_{1,t-1}^2\end{array}\right),$$
where $\beeta_t\mbox{is iid }(\bzero, \bI)$, and assuming that the components $\eta_{1t}$ and $\eta_{2t}$ are independent. Let a portfolio which is fully invested in the first asset, hence $r_t=(1,0) \by_t=y_{1t}$ for all $t$.
Denote by ${\cal F}_{1t}$ the $\sigma$-field generated by $\{y_{1u}, u\leq t\}.$ We have
$E(r_t| {\cal F}_{1,t-1})=0$ and
\begin{eqnarray*}E(r_t^2| {\cal F}_{1,t-1})&=&E(\sigma_{1t}^2| {\cal F}_{1,t-1})=\omega_1+\alpha_{11}y_{1,t-1}^2+\alpha_{12}E(y_{2,t-1}^2| {\cal F}_{1,t-1})\\
&=&\omega_1+\alpha_{11}y_{1,t-1}^2+\alpha_{12}\sigma_{2,t-1}^2 E(\eta_{2,t-1}^2| {\cal F}_{1,t-1})\\
&=&\omega_1+\alpha_{11}y_{1,t-1}^2+\alpha_{12}\sigma_{2,t-1}^2
:=\sigma_t^2.\end{eqnarray*}
It follows that $(r_t)$ satisfies the model
$r_t=\sigma_tu_t$, where
$$u_t=\frac{\sigma_{1t}}{\sigma_t}\eta_{1t}=\left(1+ \frac{\alpha_{12}\sigma_{2,t-1}^2(\eta_{2,t-1}^2-1)}{\omega_1+\alpha_{11}y_{1,t-1}^2+\alpha_{12}\sigma_{2,t-1}^2}\right)^{1/2}\eta_{1t}.$$
It is then clear that $(u_t, {\cal F}_{1,t})$ is a martingale difference but $(u_t)$ is generally not iid (except when $\alpha_{12}=0$ or $\eta_{2,t}^2$ is degenerated).
}
\end{exemple}

Even in the simple previous example,  the conditional quantile
$\mbox{VaR}_{{t_0}-1}^{*(\alpha)}(u_{t_0})$  popping up in (\ref{complicatedVaR}) cannot be explicitly computed. Whether or not this quantity could be estimated nonparametrically is beyond the scope of this paper.
Instead,
we consider a "hybrid" VaR defined by
\begin{equation}\label{hybridVaR}
\mbox{VaR}_{H,{t_0}-1}^{(\alpha)}(r_{t_0})=-\mu_{t_0}(\ba_{{t_0}-1})+\sigma_{t_0}(\ba_{{t_0}-1})\mbox{VaR}^{(\alpha)}(u)
\end{equation}
where $\mbox{VaR}^{(\alpha)}(u)$ is the marginal VaR of $u_{t}$ at level $\alpha$.
An estimator of
$\mbox{VaR}_{H,{t_0}-1}^{(\alpha)}(r_{t_0})$ is  obtained as follows: given $\ba_{{t_0}-1}=\bx$,

{\sc Step 1:} Compute the virtual historical returns $r_{t}^*(\bx)$ for $t=1, \ldots, n$.

{\sc Step 2:} Estimate $\mu_t(\bx)$ and $\sigma_t(\bx)$. Denote by $\hat{\mu}_t(\bx)$ and $\hat{\sigma}_t(\bx)$ the resulting estimators, and by
$\hat{u}_t=\{r_{t}^*(\bx)-\hat{\mu}_t(\bx)\}/\hat{\sigma}_t(\bx)$ the residuals.

{\sc Step 3:} Compute the $\alpha$-quantile $\xi_{n,\alpha}$ of  $\{\hat{u}_s, 1\leq s \leq n\}$ and define an estimator of
$\mbox{VaR}_{H,{t_0}-1}^{(\alpha)}(r_{t_0})$ as
\begin{equation}\label{Step3}
\widehat{\mbox{VaR}}_{VHS, {t_0}-1}^{(\alpha)}(r_{t_0})= -\hat{\mu}_{t_0}(\bx)-\hat{\sigma}_{t_0}(\bx)\xi_{n,\alpha}.
\end{equation}
Step 2 can be implemented by estimating a parametric model. This approach will be developed in Section \ref{secVal}.

This procedure is particularly appropriate for large portfolios, when the large dimension of the vector of underlying assets precludes--or at least formidably complicates--
estimation of multivariate volatility models.
Moreover, the following example shows that for large portfolios a univariate GARCH model is a reasonable assumption for the virtual returns.
\begin{exemple}
{\rm Suppose that $m$ is large and that the vector of log-returns is driven by a vector $\bff_t$ of $K$ factors (with $K\ll m$) as
$$\by_t=\bbeta\bff_t+\bu_t$$
where $\bbeta$ is a $m\times K$ matrix, $\mbox{var}_{t-1}(\bff_t)=\bF_t$ is a full-rank matrix  and $\mbox{var}_{t-1}(\bu_t)=\bSigma_t$. 
With a composition fixed to
$\bx$, the virtual portfolio's returns thus satisfy
$$r_t^*=\bx'\bbeta\bff_t+\bx'\bu_t.$$
Suppose that the portfolio is well-diversified, so that the components $x_i$ of $\bx$ satisfy $x_{i}=O(1/m)$ for $i=1, \ldots, m$ as $m\to \infty$.
Under appropriate assumptions 
$\mbox{var}_{t-1}(\bx'\bu_t)$
converges to 0 as $m\to \infty$.
On the other hand, $\mbox{var}_{t-1}(\bx'\bbeta\bff_t)=\bx'\bbeta \bF_t\bbeta'\bx=O_P(1)$ and does not vanish as $m$ increases under appropriate assumptions.
\footnote{For instance if the matrix $\bbeta$ does not contain too many zeroes or, more precisely, if at least one column $\bbeta_j$ of $\bbeta$
is such that $\lim \inf |\bx'\bbeta_j|>0$ as $m\to \infty$.} It follows that $r_t^*\approx \bx'\bbeta\bff_t$.
If now $K=1$ and the (real-valued) factor $f_t$ is the solution of a GARCH model, the process $\bx'\bbeta \bf_t$ will follow the same model up to a change of scale.
It is therefore natural to fit a GARCH model for the virtual returns under these assumptions when $m$ is large.
}
\end{exemple}

\section{Asymptotic properties of the VHS approach}\label{secVal}
To obtain asymptotic properties of the VHS procedure, we make the following parametric assumptions on Model (\ref{modVHS}).
For simplicity, we consider the model without conditional mean, that is $\mu_t(\bx)=0.$
For some (known) function 
$\sigma \;:\;
\mathbb{R}^{\infty}\times \Theta \rightarrow (0,\infty)$, let
\begin{equation}
\sigma_t(\bx; \btheta)=\sigma(r_{t-1}^*(\bx), r_{t-2}^*(\bx), \ldots ;
\btheta),
\end{equation}
where $\btheta_0=\btheta_0(\bx)$ is the true value of the finite dimensional parameter $\btheta$, belonging to
some compact subset $\Theta$ of $\mathbb{R}^d$. To alleviate notations, we will denote the virtual returns by $\epsilon_t:=r_{t}^*(\bx)$ and replace $\sigma_t(\bx; \btheta)$ by $\sigma_t(\btheta)$.
Model \eqref{modVHS} thus writes
\begin{equation}\label{modVHSsimplif}
\epsilon_t=\sigma_t(\btheta_0) u_t,\quad \mbox{ where for all } \btheta\in \Theta, \quad
\sigma_t(\btheta)=\sigma(\epsilon_{t-1}, \epsilon_{t-2}, \ldots ;
\btheta).
\end{equation}

Given (virtual) observations $\epsilon_1,\dots,
\epsilon_n$, and arbitrary initial values $\tilde{\epsilon}_i$ for
$i\leq 0$, we define 
$$
\tilde{\sigma}_t(\btheta)=\sigma(\epsilon_{t-1}, \epsilon_{t-2}, \ldots, \epsilon_1, \tilde{\epsilon}_0, \tilde{\epsilon}_{-1}, \ldots ;
\btheta).
$$
The Gaussian QML criterion is defined by
\begin{equation}\label{crit}
\tilde{Q}_n(\btheta)=\frac1n \sum_{t=1}^n \tilde{\ell}_t,
\qquad \tilde{\ell}_t=\tilde{\ell}_t(\btheta)= \frac{\epsilon_t^2}{\tilde{\sigma}_t^2(\btheta)}+\log \tilde{\sigma}_t^2(\btheta).
\end{equation}
Let the QML estimator of $\btheta_0$,
\begin{equation}\label{estim}
\hat{\btheta}_n= \arg\min_{\btheta\in \Theta}\tilde{Q}_n(\btheta),
\end{equation}
which has the particularity of being based on virtual returns rather than on observations.

To study the asymptotic properties of the VHS estimator,
\begin{equation}\label{Step4}
\widehat{\mbox{VaR}}_{VHS, {t_0}-1}^{(\alpha)}(r_{t_0})= -\tilde{\sigma}_{t_0}(\hat{\theta}_n)\xi_{n,\alpha},
\end{equation}
we introduce the following additional assumptions. Recall that iidness of the sequence $(u_t)$ is not a natural assumption in our framework (see Example \ref{noniid}).
Let $\bD_t(\btheta)=\sigma_t^{-1}(\btheta)\partial \sigma_t(\btheta)/\partial\btheta$, $\bD_t=\bD_t(\btheta_0) $.  Let also
${\cal F}_{t}$ the sigma-field generated by $\{u_k,k\le t\}$, and ${\cal F}_{t:t-s}$ the sigma-field generated by $\{u_k,t-s\le k\le t\}$ with $s>0$.
\begin{itemize}
\item[{\bf A1:}] The sequence $(u_t)$ is stationary, with $E|u_t|^{4+\nu}<\infty$ for some $\nu>0$, and mixing coefficients $\left\{\alpha(h)\right\}_{h\geq 0}$ satisfying, for $\epsilon\in (0,\nu)$,
$$\sum_{h=1}^{\infty}h^{r^*}\{\alpha(h)\}^{\frac{\nu-\epsilon}{4+\nu-\epsilon}}<\infty\mbox{ for some }r^*>\frac{2\kappa\{2+\nu-\epsilon\}}{\nu-\epsilon-2\kappa\{2+\nu-\epsilon\}}  \mbox{ and }\kappa\in \left(0,\frac{\nu-\epsilon}{4(2+\nu-\epsilon)}\right).$$ Suppose that $E(u_t\mid {\cal F}_{t-1})=0$ and $E(u^2_t\mid {\cal F}_{t-1})=1$.
Let
$\xi_{\alpha}$ the $\alpha$-quantile of $u_t$.
Assume that the conditional distribution of $u_t$ given ${\cal F}_{t-1}$ has a density $f_{t-1}$ 
    such that $f_{t-1}(\xi_{\alpha})>0$ a.s. and $E\sup_{\xi\in V(\xi_{\alpha})}f^4_{t-1}(\xi)<\infty$ for some neighborhood $V(\xi_{\alpha})$ of $\xi_{\alpha}$. Assume also that this density is continuous at $\xi_{\alpha}$ uniformly in ${\cal F}_{t-1}$, in the sense that for sufficiently small $\varepsilon> 0$, there exists a stationary and ergodic sequence $(K_t)$ such that $K_{t-1}\in {\cal F}_{t-1}$ and
$$\sup_{x\in[\xi_{\alpha}-\varepsilon,\xi_{\alpha}+\varepsilon]}\left|f_{t-1}(x)-f_{t-1}(\xi_{\alpha})\right|\leq K_{t-1}\varepsilon$$
with $EK^4_{t}<\infty$ a.s.
\item[{\bf A2:}]
 $(\epsilon_t)$ is a strictly stationary and ergodic solution of (\ref{modVHSsimplif}), and there exists $s_0>0$ such that $E|\epsilon_1|^{s_0}<\infty$.
\item[{\bf A3:}] There exists a sequence $\bD_{t,T_n}$ such that
$\bD_t=\bD_{t,T_n}+o_P(1)\mbox{ as } n\to\infty$,
where $T_n\to\infty$ and $T_n=O(n^{\kappa})$ 
(with $\kappa$ defined in {\bf A1}), 
$\bD_{t,T_n}\in {\cal F}_{t-1:t-T_n}$
and for any $r\geq 0$
$$E\|\bD_{t}\|^r<\infty,\qquad \sup_{n\geq 1}E\|\bD_{t,T_n}\|^r<\infty. 
$$
\item[{\bf A4:}]
For
some $\underline{\omega}>0$, almost surely, $\sigma_t(\btheta)$ and $\tilde{\sigma}_t(\btheta)$ belong to $(\underline{\omega}, \infty]$
for any $\btheta\in \Theta$ and any $t\ge 1$. For $\btheta_1, \btheta_2\in \Theta$, we have
$\sigma_t(\btheta_1)=\sigma_t(\btheta_2)\; a.s.$ if and only if
$\btheta_1=\btheta_2$. Moreover, for any $\bx\in \mathbb{R}^m$, $\bx'\bD_t(\btheta_0)=0$ a.s. entails $\bx=0$.
\item[{\bf A5:}]
There exist a random variable $C$ which is measurable with respect to $\{\epsilon_t, t<0\}$ and a constant $\rho\in(0,1)$ such that $\sup_{\btheta\in\Theta}|\sigma_t(\btheta)-\widetilde{\sigma}_t(\btheta)|\leq C\rho^t$.
\item[{\bf A6:}] The function $\btheta \mapsto \sigma (x_1, x_2, \ldots ; \btheta)$ has continuous second-order derivatives, and
$$\sup_{\btheta\in\Theta}\left\|\frac{\partial
\sigma_t(\btheta)}{\partial \btheta}
-\frac{\partial\widetilde{\sigma}_t(\btheta)}{\partial \btheta} \right\|
+\left\|\frac{\partial^2
\sigma_t(\btheta)}{\partial \btheta\partial \btheta'}
-\frac{\partial^2\widetilde{\sigma}_t(\btheta)}{\partial \btheta\partial \btheta'} \right\|
\leq
C\rho^t,$$
%
where $C$ and $\rho$ are as in {\bf A5}.
\item[{\bf A7:}] There exists a neighborhood $V(\btheta_0)$ of $\btheta_0$ and $\tau>0$ such that
$$\sup_{\btheta \in V(\btheta_0)} \left\|\frac1{\sigma_t(\btheta)}\frac{\partial \sigma_t(\btheta)}{\partial\btheta}\right\|^4, \quad
\sup_{\btheta \in V(\btheta_0)} \left\|\frac1{\sigma_t(\btheta)}\frac{\partial^2 \sigma_t(\btheta)}{\partial\btheta\partial\btheta'}\right\|^2, \quad
\sup_{\btheta \in V(\btheta_0)} \left|\frac{\sigma_t(\btheta_0)}{\sigma_t(\btheta)}\right|^{\frac{2(4+\nu)(1+\tau)}{2+\nu}},$$
have finite expectations, where $\nu$ is as in {\bf A1}.
\end{itemize}

Assumptions {\bf A2}, {\bf A4} and {\bf A5}, together with $E(u^2_t\mid {\cal F}_{t-1})=1$, are sufficient to prove the strong consistency of the QML estimator $\hat{\btheta}_n$.
Assumption {\bf A6} allows to show that the initial values $\tilde{\epsilon}_i$ do not matter for the asymptotic normality of  $\hat{\btheta}_n$.
Assumptions {\bf A1}, {\bf A3} and {\bf A7} are used to apply a CLT for a mixing triangular array based on the approximation of $\bD_t$ by $\bD_{t,T_n}$.

For particular volatility models, some of the assumptions can be simplified as the following lemma shows.
\begin{lem}\label{exempleGARCH}
For the standard GARCH(1,1) model
\begin{equation}\label{GARCH11}
\epsilon_t=\sigma_t u_t,\quad
\sigma_t^2=\omega_0+\alpha_0\epsilon_{t-1}^2+ \beta_0\sigma_{t-1}^2,\qquad
\omega_0>0, \alpha_0> 0, \beta_0>0,
\end{equation}
where $(u_t)$ satisfies {\bf A1}, 
Assumptions {\bf A2-A7} reduce to:
i) $E\log (\alpha_0u_t^2+\beta_0)<0$; ii) $u_t^2$ has a non-degenerate distribution; 
iii) $\Theta=\{(\omega, \alpha, \beta)\}$ is a compact subset of $(0,\infty)^3$ such that, for all $\theta\in \Theta$,  $\omega>\underline{\omega}$ for some $\underline{\omega}>0$ and $\beta<1$.
iv) $E|\epsilon_t|^{s_0}<\infty$ for $s_0>0$, where $(\epsilon_t)$ is the strictly stationary solution implied by i).
\end{lem}

We are now in a position to state our main result.
\begin{theo}\label{hard}
Assume $\xi_{\alpha}<0$. Let {\bf A1-A7} hold. Then $\hat{\btheta}_n\to\btheta_0$ a.s. as $n\to \infty$, and
\begin{eqnarray*}
\left(\begin{array}{c}\sqrt{n}\left(\hat{\btheta}_n-\btheta_0\right) \\
\sqrt{n}(\xi_{\alpha}-\xi_{n,
\alpha})
\end{array}\right)
&\stackrel{{\cal L}}{\to}&{\cal N}(0,\bSigma_{\alpha}),
\qquad \bSigma_{\alpha}=\left(\begin{array}{cc}
\bJ^{-1}\bS^{11}\bJ^{-1}& \bLambda_{\alpha}\\ \bLambda_{\alpha}'    &\zeta_{\alpha}
\end{array}\right),
\end{eqnarray*}
where   $\bJ=E(\bD_t\bD_t')$ with $\bD_t=\bD_t(\btheta_0)$, and
\begin{eqnarray*}
\bLambda_{\alpha}&=&
\frac{\xi_{\alpha}}{4E[f_{t-1}(\xi_{\alpha})]}\bJ^{-1}\bS^{11}\bJ^{-1}\bPsi_{\alpha}+\frac1{2E[f_{t-1}(\xi_{\alpha})]}\bJ^{-1}\bS_{\alpha}^{12}, \qquad \bPsi_{\alpha}=E[f_{t-1}(\xi_{\alpha})\bD_t],\\
\zeta_{\alpha}&=&\frac1{\{E[f_{t-1}(\xi_{\alpha})]\}^2}\left(\frac14\xi_{\alpha}^2\bPsi_{\alpha}'\bJ^{-1}\bS^{11}\bJ^{-1}\bPsi_{\alpha}+ \xi_{\alpha}\bPsi_{\alpha}'\bJ^{-1}\bS_{\alpha}^{12}+\bS_{\alpha}^{22}\right),\\
\bS^{11}&=&E\left[\left(u_t^2-1\right)^2\bD_t\bD_t'\right], \qquad
\bS_{\alpha}^{22}=\sum_{h=-\infty}^{\infty}\mbox{cov}\left(\mathbf{1}_{\{u_t<\xi_{\alpha}\}},\mathbf{1}_{\{u_{t-h}<\xi_{\alpha}\}}\right),\\
\bS_{\alpha}^{12}&=&\bS_{\alpha}^{21 '}=\sum_{h=0}^{\infty}\mbox{cov}[(u_t^2-1)\bD_t,\mathbf{1}_{\{u_{t+h}<\xi_{\alpha}\}}].
\end{eqnarray*}
\end{theo}
\begin{rem} When the errors $u_t$ are iid, the expression of the asymptotic covariance matrix simplifies considerably. More precisely, we have
\begin{eqnarray*}
\bS^{11}&=&\left(\kappa_4-1\right)\bJ, \qquad
\bS_{\alpha}^{22}=\alpha(1-\alpha),\quad
\bS_{\alpha}^{12}=\bS_{\alpha}^{21 '}=[E\left(u_t^2\mathbf{1}_{\{u_{t}<\xi_{\alpha}\}}\right)-\alpha]E(\bD_t):= p_{\alpha}E(\bD_t),\\
\bLambda_{\alpha}&=&
\left(\frac{\xi_{\alpha}(\kappa_4-1)}{4}+\frac{p_{\alpha}}{2f(\xi_{\alpha})}\right)  \bJ^{-1}\bOmega, \qquad \bOmega=E(\bD_t),\\
\zeta_{\alpha}&=&\frac{\kappa_4-1}4\xi_{\alpha}^2\bOmega'\bJ^{-1}\bOmega+\frac{\xi_{\alpha}p_{\alpha}}{f(\xi_{\alpha})}\bOmega'\bJ^{-1}\bOmega+\frac{\alpha(1-\alpha)}{f^2(\xi_{\alpha})}=
\frac{\kappa_4-1}4\xi_{\alpha}^2+\frac{\xi_{\alpha}p_{\alpha}}{f(\xi_{\alpha})}+\frac{\alpha(1-\alpha)}{f^2(\xi_{\alpha})}.
\end{eqnarray*}
where $f$ denotes the density of $u_t$, and $\kappa_4=Eu_t^4$.
For the last equality, we used the relation
$\bOmega'\bJ^{-1}\bOmega=1$ (see (8) in Francq and Zakoïan (2013)). Hence, when $(u_t)$ is iid we have
$$\bSigma_{\alpha}=\left(\begin{array}{cc}
\left(\kappa_4-1\right)\bJ^{-1}& \left(\frac{\xi_{\alpha}(\kappa_4-1)}{4}+\frac{p_{\alpha}}{2f(\xi_{\alpha})}\right)  \bJ^{-1}\bOmega\\
\left(\frac{\xi_{\alpha}(\kappa_4-1)}{4}+\frac{p_{\alpha}}{2f(\xi_{\alpha})}\right)\bOmega'\bJ^{-1}    &\frac{\kappa_4-1}4\xi_{\alpha}^2+\frac{\xi_{\alpha}p_{\alpha}}{f(\xi_{\alpha})}+\frac{\alpha(1-\alpha)}{f^2(\xi_{\alpha})}
\end{array}\right).$$
In particular, the asymptotic variance of $\sqrt{n}(\xi_{\alpha}-\xi_{n,\alpha})$ only depends on the errors distribution, not on the volatility parameter $\btheta_0$. Nevertheless, the estimation of $\btheta_0$ affects the asymptotic accuracy, since the
asymptotic variance would be equal to $\frac{\alpha(1-\alpha)}{f^2(\xi_{\alpha})}$ if the $u_t$'s were observed.
\end{rem}
\begin{rem} Consistent estimation of ${\bSigma}_{\alpha}$ is crucial for evaluating the estimation risk.
Some matrices involved in the asymptotic covariance matrix have the form of expectations, and can therefore be straightforwardly estimated
by their empirical counterpart. This is the case of $\bJ$ and $\bS^{11}$. To estimate $\bS^{22}_{\alpha}$ and $\bS^{12}_{\alpha}$, a classical HAC (heteroskedasticity and autocorrelation consistent) estimator can be used
(see for instance Andrews (1991), Newey and West (1987)).
Estimation of the matrix $\bPsi_{\alpha}$ is more tricky. We propose two approaches: i) noting that
\begin{eqnarray*}
\bPsi_{\alpha}&=&\lim_{h\to 0} \frac 1h E\left[\{P_{t-1}(u_t<\xi_{\alpha}+h)-P_{t-1}(u_t<\xi_{\alpha})\}\bD_t\right]\\
&=&\lim_{h\to 0} \frac 1h E\left[E_{t-1}\{\mathbf{1}_{\{u_t<\xi_{\alpha}+h\}}-\mathbf{1}_{\{u_t<\xi_{\alpha}\}}\}\bD_t\right]=
\lim_{h\to 0} \frac 1h E\left[(\mathbf{1}_{\{u_t<\xi_{\alpha}+h\}}-\mathbf{1}_{\{u_t<\xi_{\alpha}\}})\bD_t\right],
\end{eqnarray*}
a natural estimator for $\bPsi_{\alpha}$ is
\begin{eqnarray*}
\widehat{\bPsi}_{\alpha}&=&
 \frac1{nh_n} \sum_{t=1}^n(\mathbf{1}_{\{\hat{u}_t<\xi_{n,\alpha}+h_n\}}-\mathbf{1}_{\{\hat{u}_t<\xi_{n,\alpha}\}})\hat{\bD}_t,
\end{eqnarray*}
where $\hat{\bD}_t=\tilde{\sigma}_t^{-1}(\hat{\btheta}_n)\partial \tilde{\sigma}_t(\hat{\btheta}_n)/\partial\btheta$
and $h_n$ is a bandwidth parameter; ii) alternatively, one may approximate $f_{t-1}$ by the density of $u_t$ conditional on $u_{t-1}$ (instead of the infinite past).
A standard kernel estimator for this density at point $\xi_{\alpha}$ is
$$\hat{f}_{t-1}(\xi_{\alpha})= \frac{\sum_{s=1}^nK_{h_2}(\hat{u}_{t-1}-\hat{u}_{s-1})K_{h_1}(\xi_{\alpha}-\hat{u}_{s})}{\sum_{s=1}^nK_{h_2}(\hat{u}_{t-1}-\hat{u}_{s-1})},$$
where $h_1, h_2$ are bandwidths, 
$K_h(x)=h^{-1}K(x/h)$ for any bandwidth $h$, and $K$ is a bounded symmetric kernel function (see e.g. Hansen (2004)).
Matrix $\bPsi_{\alpha}$ can thus be estimated by
\begin{eqnarray*}
\widehat{{\bPsi}}^{\circ}_{\alpha}&=&\frac 1n\sum_{t=1}^n \hat{f}_{t-1}(\xi_{\alpha})\hat{\bD}_t.
\end{eqnarray*}
The asymptotic properties under Assumptions {\bf A1-A7} of both estimators, $\widehat{\bPsi}_{\alpha}$ and $\widehat{{\bPsi}}^{\circ}_{\alpha}$, are left for further research.
\end{rem}
 We have the Taylor expansion
$$F_{t_0}(\hat{\btheta}_n, \xi_{n,\alpha}):=\tilde{\sigma}_{t_0}(\hat{\btheta}_n)\xi_{n,\alpha}=F_{t_0}({\btheta}_0, \xi_{\alpha})+ \frac{\partial}{\partial (\btheta', \xi)}{F}_{t_0}({\btheta}_n^*, \xi_{n}^*)
\sqrt{n}\left(\begin{array}{c}\hat{\btheta}_n-\btheta_0 \\
\xi_{\alpha}-\xi_{n,
\alpha}
\end{array}\right),$$
where $({\btheta}_n^*, \xi_n^*)$ is between $({\btheta}_n, \xi_{n,\alpha})$ and $(\btheta_0, \xi_{\alpha})$.
 Given a fixed information set
 $I_{t_0-1}$, it can be shown using {\bf A5-A6} that the observations available at time $t_0-1$ have no effect on the asymptotic distribution of Theorem \ref{hard}. Thus, an approximate {\it conditional} $(1-\alpha_0)\%$
CI for the hybrid VaR 
 \eqref{hybridVaR} has  bounds given by
\begin{equation}
\label{CIuniv}
\widehat{\mbox{VaR}}_{VHS, t_0-1}^{(\alpha)}(r_{t_0})
\pm\frac1{\sqrt{n}}
\Phi^{-1}(1-\alpha_0/2)
\left\{
\bdelta_{{t_0}-1}'\widehat{\bSigma}_{\alpha} \bdelta_{{t_0}-1}
\right\}^{1/2},
\end{equation}
where $\widehat{\mbox{VaR}}_{VHS, t_0-1}^{(\alpha)}(r_{t_0})$
is defined in \eqref{Step4}, $\widehat{\bSigma}_{\alpha}$ is a consistent estimator of
${\bSigma}_{\alpha}$ and
\begin{equation*}
\bdelta_{{t_0}-1}'=
\left(\frac{\partial
\tilde{\sigma}_{t_0}(\hat{\btheta}_n)}{\partial \btheta}\xi_{n,\alpha}\qquad
\tilde{\sigma}_{t_0}(\hat{\btheta}_n)\right).
\end{equation*}
In other words,
$$\lim_{n\to \infty}
P_{t_0-1}\left[\left|{\mbox{VaR}}_{H, t_0-1}^{(\alpha)}(r_{t_0})-
\widehat{\mbox{VaR}}_{VHS, t_0-1}^{(\alpha)}(r_{t_0})\right|
\le \frac1{\sqrt{n}}
\Phi^{-1}(1-\alpha_0/2)
\left\{
\bdelta_{{t_0}-1}'\widehat{\bSigma}_{\alpha} \bdelta_{{t_0}-1}
\right\}^{1/2}\right]= 1-\alpha_0.
$$
\begin{rem} It is worth noting that the previous CI has to be understood conditionally on $I_{t_0-1}$ for fixed $t_0$.
In particular, it cannot be directly applied when  $t_0-1=n$.
Indeed, the bounds in (\ref{CIuniv}) are no longer random  in this case.
\footnote{except if one assumes, as in
Gong, Li and Peng (2010), that the number of conditioning values is finite as in the ARCH$(q)$ case.}
See Beutner, Heinemann and Smeekes (2019) for an asymptotic justification of this type of CI for conditional objects.
\end{rem}

\section{Numerical illustrations}\label{secillus}

An alternative to the univariate approaches is a multivariate strategy, requiring a dynamic model for the vector of
risk factors $\by_t$. We describe in Appendix \ref{secMul} two multivariate approaches -- the spherical method and the
Filtered Historical Simulation (FHS) method. Such methods might perform better than univariate ones because
they incorporate information stemming from individual returns rather than an aggregate information stemming from portfolio returns.
  However, at least for large portfolios,  multivariate approaches can be very challenging, when at all possible.



The first section is devoted to numerical comparisons of the different methods.
We first study the reliability of the naive approach in a static framework where the returns are iid and the portfolio is crystallized.
In a dynamic framework, we then compare the performance of the univariate and multivariate approaches when the dynamic multivariate model is well specified and the dimension is small.
Next, we consider
misspecified GARCH models and, possibly, higher dimensions.
The second section concerns
real data examples based on large sets of US
stocks.\footnote{ The code and data used in the paper are available on the web site\\
\url{http://perso.univ-lille3.fr/~cfrancq/Christian-Francq/VaRPortfolio.html}
}

\subsection{
Monte-Carlo experiments}

The theoretical conditional VaR in (\ref{truevar}) depends on the information set $I_{t-1}$.
In our first two sets of experiments, the vector of individual returns will be simulated using multivariate GARCH
and
the different estimators will be compared to the same target VaR obtained by
taking the full information set (i.e. including the returns of the individual assets, not only the returns of the portfolio).
In the third set of experiments, we will estimate a misspecified multivariate model for which the true conditional VaR is no longer available.

\subsubsection{
Static model and crystallized portfolio
}\label{appendE}
\begin{figure}
\vspace*{10cm} \protect \includegraphics{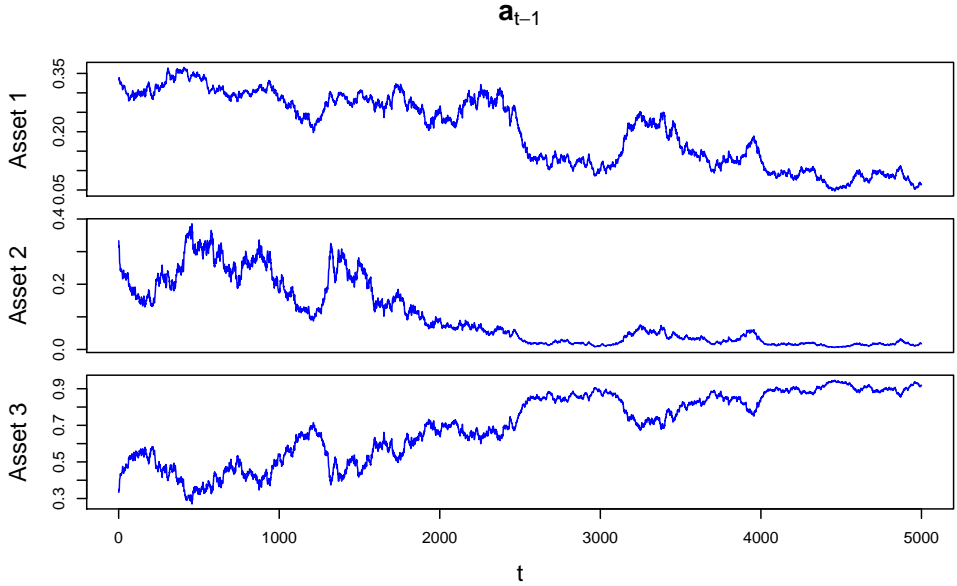}
\vspace{-2cm}
\caption{\label{Composition1}{\small Time-varying composition of the simulated crystallized portfolio.}}
\end{figure}

For a simple illustration of Lemma \ref{paire}, we consider a crystallized
equally weighted portfolio of 3 assets (of initial price $p_{i0}=1000$):
$V_t=\sum_{i=1}^3 p_{it}$ and $\mu_{i,t}=1$ for $i=1,2,3$.
Thus, the return portfolio composition is time varying, with coefficients
$\mathbf{a}_{t-1}=(a_{1,t-1}, a_{2,t-1}, a_{3,t-1})'$ and
$a_{i,t-1}=p_{i,t-1}/\sum_{j=1}^3 p_{j,t-1}$.
Assume that the vector of the log-returns is iid, Gaussian, centered, with variance
$\mbox{Var}(\by_t)=\bSigma^2=\bD\bR\bD,$
with
$$\bD=\left(\begin{array}{ccc}0.01&0&0\\0&0.02&0\\
0&0&0.04\\\end{array}\right),\quad \bR=\left(\begin{array}{ccc}1&-0.855&0.855\\-0.855&1&-0.810
\\
0.855&-0.810&1\end{array}\right).$$
The composition $\ba_{t-1}$ of the portfolio is plotted in Figure \ref{Composition1}. As we have seen in Section \ref{secInval}, this
vector is non stationary. 
More precisely, as discussed in Section \ref{secInval}, with increasing probability the composition $\ba_{t-1}$
of the portfolio is arbitrarily close to one of the three single-asset portfolios
$(1,0, 0), (0,1,0)$ and $(0,0,1)$. Experiments conducted with different parameters (in particular different correlation matrices) led to the same conclusion.

\begin{figure}
\begin{center}
\vspace*{9cm} \protect \includegraphics{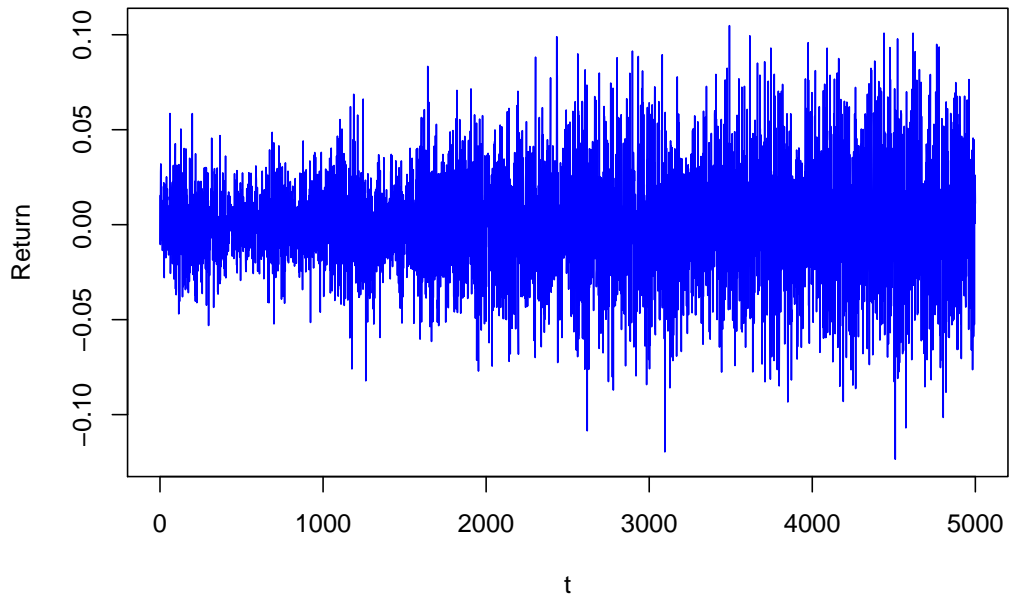}
\end{center}\vspace{-3cm}
\caption{\label{epsilon1}{\small Returns of the simulated crystallized portfolio.}}
\end{figure}

It is thus non surprising to see that the univariate return series $r_t$ plotted in Figure \ref{epsilon1}
exhibits some nonstationarity features, in particular unconditional heteroscedasticity. The increased variance in the second part of the sample reflects the
fact that the portfolio tends to be less and less diversified (see Figure \ref{Composition1}).

\begin{figure}
\begin{center}
\vspace*{7cm} \protect \includegraphics{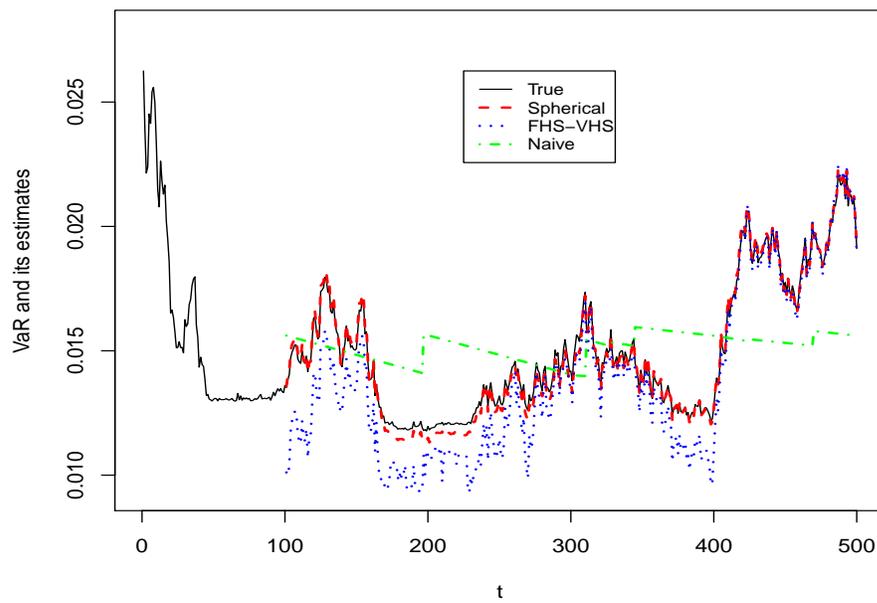}
\end{center}\vspace{1cm}
\caption{\label{VaRestimates1}{\small True and estimated VaRs of the crystallized portfolio.}}
\end{figure}

When $\bSigma_t$ and $\bm_t$ are constant, the FHS estimator in \eqref{FHS} reduces to the opposite of the quantile of the portfolio's virtual returns:
$\widehat{\mbox{VaR}}^{(\alpha)}_{FHS,t-1}(r_t) %
:=-q_{\alpha}\left(\{\ba_{t-1}'\by_1,\dots,\ba_{t-1}'\by_{t-1}\}\right)$ and this estimator coincides with the VHS estimator.
These empirical quantiles were computed starting from $t=100$.
The naive estimator is the opposite of the quantile of the portfolio's returns:
$\widehat{\mbox{VaR}}^{(\alpha)}_{N,t-1}(r_t) %
=-q_{\alpha}\left(\{\ba_{1}'\by_1,\dots,\ba_{t-1}'\by_{t-1}\}\right)$.
The spherical method, based on the estimation of $\bSigma$, was computed on the same range of observations.
In view of (\ref{VaRunderS}), we have
$\widehat{\mbox{VaR}}^{(\alpha)}_{S,t-1}(r_t)
=\sqrt{\ba_{t-1}'\widehat{\bSigma}_{t-1}\ba_{t-1}}q_{1-2\alpha}\left(\{\widehat{\bSigma}_{t-1}^{-1}\by_1,\dots,\widehat{\bSigma}_{t-1}^{-1}\by_{t-1}\}\right)$, where
$\widehat{\bSigma}_{t-1}$ is the empirical covariance matrix of the observations $\by_1,\ldots,\by_{t-1}$.
Figure \ref{VaRestimates1} displays the sample paths of the true conditional VaR as well as the estimated VaRs.
It can be seen
that the spherical method converges faster to the true value than the FHS=VHS method.
Unsurprisingly, the univariate naive method fails to converge to the theoretical conditional VaR based on the full information set.
Now we turn to a less artificial setting.

\subsubsection{Well-specified multivariate GARCH models}\label{secestmul}
In this section, we simulate
the process of log-returns $\by_t=\bSigma_t\beeta_t$ from
the corrected Dynamic Conditional Correlation (cDCC) GARCH model of Aielli (2013). 
For the multivariate approaches, we estimate the same cDCC-GARCH(1,1) model. For the univariate approaches we estimate
GARCH(1,1) models, which are generally misspecified (see Example \ref{noniid}).

We consider the minimum variance portfolio variance given by
\begin{equation}
\label{portopt}
r_t=\by_t'{\ba}_{t-1},\quad \ba_{t-1}=\frac{\bSigma^{-2}_t\be}{\be'\bSigma^{-2}_t\be}, \quad \mbox{where} \quad \be=(1, \ldots, 1)'\in \mathbb{R}^m.
\end{equation}
 We simulated $N$ independent trajectories of length $n$ for the cDCC-GARCH(1,1) model. On each simulation, the first $n_1$ observations are used (i) to obtain an estimator $\widehat{\bvartheta}_{n_1}$ of the parameters involved in
  $\bSigma_t$ by the three-step estimator defined in Appendix C of Francq and Zakoian (2018), and (ii) to estimate the quantiles required for the VaR estimator.
  On the last $n-n_1$ simulations, {\it i.e.} for $t=n_1+1,\dots,n$, we compared the theoretical VaR of the portfolio
with the three estimates obtained from the spherical, FHS and VHS methods.
We considered portfolios of $m=2$ assets. The different designs, displayed in
Appendix \ref{appendDCC}, correspond to spherical (designs A-H) or non spherical (designs A$^*$-H$^*$)
distributions.



We took $N=100$ independent replications, and $n-n_1=1000$ out-of-sample predictions for each simulation. In each design, we then compared the corresponding $100, 000$ theoretical values of the conditional VaR
with their estimates 
obtained by the spherical, FHS and VHS methods.
Denote by $\mathrm{MSE}_S$, $\mathrm{MSE}_{FHS}$ and $\mathrm{MSE}_{VHS}$ the mean square errors (MSE) of prediction of the three methods.
Table~\ref{ARE} displays the relative efficiency (RE) of the spherical method with respect to the FHS and the VHS methods, as measured by the ratios $\mathrm{MSE}_{FHS}/\mathrm{MSE}_S$ and $\mathrm{MSE}_{VHS}/\mathrm{MSE}_S$.
It should be underlined that all MSE's are computed with respect to the full information VaR's, which a priori favours the multivariate methods.
Let us first briefly compare the two multivariate approaches:  in designs A-H (with spherical distributions) 
the spherical method is as expected more efficient than the FHS method (for Designs C and D, the spherical method can be twice more efficient than the other multivariate method).
On the contrary, the bottom panel of
Table~\ref{ARE} reveals that, when the density is  strongly asymmetric, the FHS method can be much more efficient than the spherical method.

One question of interest is whether the VHS estimator (targeted to estimate the hybrid VaR) can be used to approximate this full-information VaR.
Let us first compare the VHS method to the multivariate approaches in the spherical case.
The univariate VHS method is apparently dominated by the multivariate methods in designs A-H
but one has to recall that the reference VaR to which the methods are compared is designed for the multivariate framework.
Better results are obtained for the VHS method in designs E-H (identically distributed returns) as opposed to A-D, 
and for designs $\{$A,B,E,F$\}$ as opposed to $\{$C,D,G,H$\}$ (independent returns). This can be intuitively explained as follows.
Univariate methods are expected to behave better when the trajectories of the underlying returns are close, which is the case when the two assets
have similar dynamic models and are strongly dependent. It is thus non surprising to note that the worst results for the VHS (by comparison with the multivariate methods)
occur for designs C-D (independent returns with very different dynamics), and the best results occur with designs E-F (dependent identically distributed returns).

In the case of non-spherical distributions (designs A*-H*)
the same conclusions hold:
the more dependent the assets and the closer their dynamics, the better results for the VHS. 
This method clearly outperforms the spherical approach in designs E*-H*.
Unexpectedly, 
the results reveal that the univariate VHS method can even outperform the FHS approach  (design E* with $n_1=1000$).


From these experiments, it appears that the accuracy of the approximation provided by the VHS approach is very dependent from the model parameters.
However, these first two examples clearly favour the multivariate methods 
by assuming that the dynamic model is 
well specified.
In the next example, we consider a data generating process that does not belong to the GARCH class, and we will compare the different methods using backtests.

\renewcommand{\baselinestretch}{1}\small\normalsize
\begin{center}
\begin{table}
\caption{\label{ARE}\footnotesize
Relative efficiency  of the Spherical method with respect to the FHS method ($\mathrm{MSE}_{FHS}/\mathrm{MSE}_{S}$) and with respect to the VHS method ($\mathrm{MSE}_{VHS}/\mathrm{MSE}_{S}$) for predicting the full-information VaR.
Designs A-H correspond to spherical distributions, Designs A$^*$-H$^*$ correspond to non-spherical distributions (see Appendix \ref{appendDCC}).
 The number of independent replications of each design  is $100$ and, for each simulation, $n_1$ observations were used for estimation and
$1000$ out-of-sample predictions were computed. }
\begin{center}
\begin{tabular}{lllcccccccc}
\hline\hline
$n_1$  &  $\alpha$   & &A&B&C&D&E&F&G&H                                        \\
500    &     1\% &$\mathrm{MSE}_{FHS}/\mathrm{MSE}_{S}$&1.12 &1.11& 2.57& 2.35 &1.08 &1.17 &1.23 &1.42           \\
       &         &$\mathrm{MSE}_{VHS}/\mathrm{MSE}_{S}$&53.9&  16.9& 188.&  82.0&   1.63 &  1.53&   2.42&   2.18 \\
       & 5\%     &$\mathrm{MSE}_{FHS}/\mathrm{MSE}_{S}$&1.21 &1.03 &1.81 &1.40 &1.18 &1.12 &1.12 &1.19           \\
       &         &$\mathrm{MSE}_{VHS}/\mathrm{MSE}_{S}$& 28.7& 9.76& 130.& 74.6& 1.69& 1.56& 2.31& 1.99          \\
1000   & 1\%     &$\mathrm{MSE}_{FHS}/\mathrm{MSE}_{S}$&1.30 &1.11 &2.35 &1.62 &1.53 &1.51 &1.57 &1.36           \\
       &         &$\mathrm{MSE}_{VHS}/\mathrm{MSE}_{S}$&91.6  &23.4& 303.  &79.8  & 1.93  & 2.53   &4.43   &2.23 \\
       & 5\%     &$\mathrm{MSE}_{FHS}/\mathrm{MSE}_{S}$& 1.14& 1.03& 2.07& 1.00 &1.25 &1.08 &1.33 &1.01          \\
       &         &$\mathrm{MSE}_{VHS}/\mathrm{MSE}_{S}$& 55.4 & 15.7 & 267. & 82.5 &  1.75  & 2.44  & 4.14& 2.01 \\
                     &&& A$^*$&B$^*$&C$^*$&D$^*$&E$^*$&F$^*$&G$^*$&H$^*$       \\
500    & 1\%     &$\mathrm{MSE}_{FHS}/\mathrm{MSE}_{S}$& 0.08 &0.06 &0.03 &0.04 &0.12 &0.10 &0.10 &0.12          \\
       &         &$\mathrm{MSE}_{VHS}/\mathrm{MSE}_{S}$&2.70& 4.45& 2.78& 3.32& 0.18& 0.14& 0.16& 0.18           \\
       & 5\%     &$\mathrm{MSE}_{FHS}/\mathrm{MSE}_{S}$& 0.32 &0.32 &0.13 &0.25 &0.45 &0.52 &0.46 &0.53          \\
       &         &$\mathrm{MSE}_{VHS}/\mathrm{MSE}_{S}$& 5.3 &11.0& 10.2& 14.9 & 0.65&  0.66&  0.71 & 0.73       \\
1000   & 1\%     &$\mathrm{MSE}_{FHS}/\mathrm{MSE}_{S}$& 0.08 &0.03 &0.02 & 0.02 &0.06 &0.03 &0.03 &0.04         \\
       &         &$\mathrm{MSE}_{VHS}/\mathrm{MSE}_{S}$&  2.20& 2.43 &2.31& 1.67 &0.05 &0.04 &0.07 &0.06         \\
       & 5\%     &$\mathrm{MSE}_{FHS}/\mathrm{MSE}_{S}$& 0.34& 0.19& 0.09 &0.11 &0.30& 0.24 &0.21 &0.29          \\
       &         &$\mathrm{MSE}_{VHS}/\mathrm{MSE}_{S}$& 3.78 & 6.68 &10.2 & 8.72 & 0.26 & 0.35 & 0.59 & 0.44    \\
\\
\hline
\end{tabular}
\end{center}
\end{table}
\end{center}

\renewcommand{\baselinestretch}{1.4}\small\normalsize

\subsubsection{
Misspecified GARCH models}
\begin{center}
\begin{figure}
\vspace*{8.cm}
\includegraphics{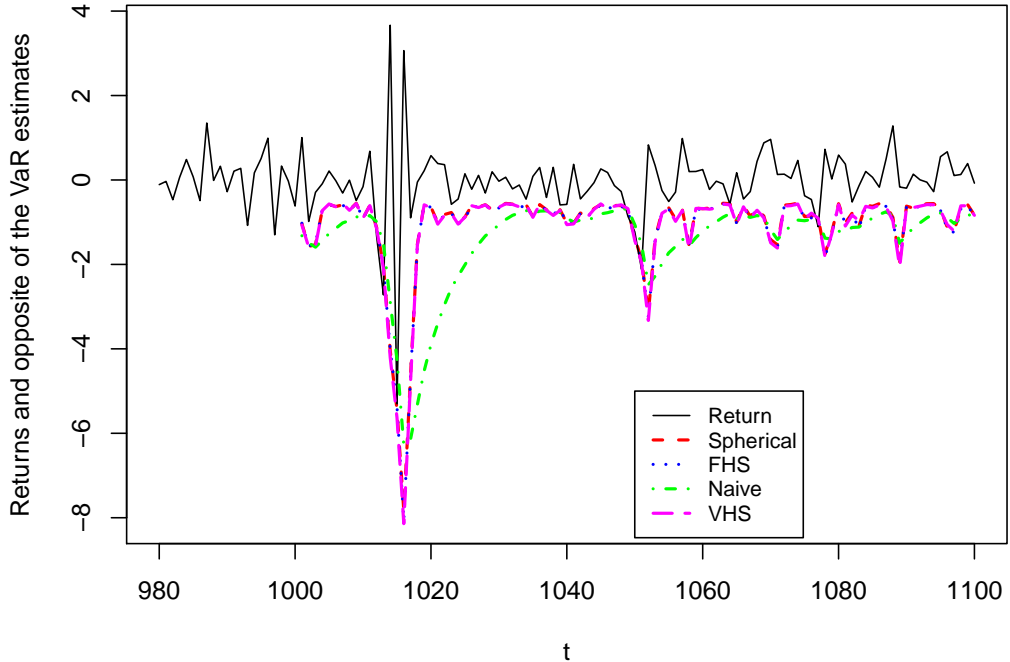}
\vspace{0.9cm}
\caption{\label{4VaR} {\small
Comparison of 4 VaRs at the horizon 1 for a crystallized portfolio of $m=4$ simulated assets.}}
\end{figure}
\end{center}

\begin{center}
\begin{figure}
\vspace*{9.cm}
\includegraphics{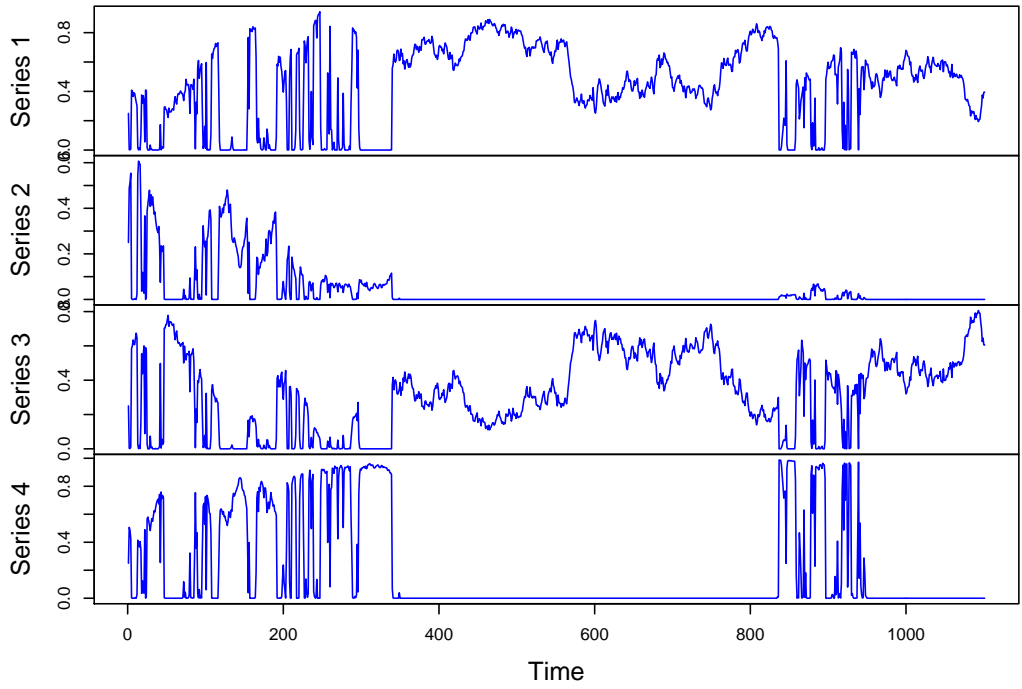}
\vspace{0.9cm}
\caption{\label{compo} {\small
Time-varying composition of the return of the crystallized portfolio as function of the returns of the individual assets.}}
\end{figure}
\end{center}

We simulated $m$-multivariate factor models, with two GARCH factors of the form
$$f_{1t}=\sigma_{1t}\eta_{1t}, \quad f_{2t}=\sigma_{2t}\eta_{2t},$$
where $(\eta_{1t})_t$ and $(\eta_{2t})_t$ are two independent sequences of iid ${\cal N}(0,1)$-distributed random variables. The volatilities follow standard GARCH$(1,1)$ equations of the form
$$\sigma_{it}^2=\omega_i+\alpha_if_{i,t-1}^2+\beta_i\sigma_{i-1,t}^2.$$
We took $(\omega_1,\alpha_1,\beta_1)=(1,0.09,0.87)$ and $(\omega_2,\alpha_2,\beta_2)=(0.1,0.7,0.01)$, so that the dynamics of the two factors be quite distinct.
The even and odd components of our simulated factor model are respectively of the form
$$y_{2k,t}=f_{2t}+e_{2k,t},\qquad y_{2k+1,t}=f_{1t}+e_{2k+1,t},$$
where $(e_{k,t})_t$, for $k=1,\dots,m$, are idiosyncratic independent iid noises with law ${\cal N}(0,0.1^2)$.
To obtain a graphical comparison  of the VaR estimates, we first simulated a trajectory of size $1, 100$ of the factor model with $m=4$. A crystallized portfolio of composition $(1/m,\dots,1/m)$ at time $t=1$ has been considered.
For the multivariate approaches we estimated cDCC-GARCH(1,1) models, while for the univariate approaches we estimated GARCH(1,1) models. The four competing estimators of the 5\% $\mbox{VaR}_{t-1}$ at time $t=1001$ were estimated on the basis on the first $1, 000$ simulated values $\by_1,\dots,\by_{t-1}$.
Then, VaR at time $t=1, 002$ was estimated based on the past $1, 000$ simulations  $\by_2,\dots,\by_{t-1}$. We continued this way until we obtained the last VaR estimations  at time $t=1,100$. Figure \ref{4VaR} shows that the estimates obtained by the Spherical, FHS and VHS methods are very close (actually, they are not distinguishable on the figure), whereas the estimates obtained by the naive method behave differently. This can be explained by the fact that the portfolio is crystallized but not static. In other words, even if the portfolio is constituted of an equal quantity of the $m$ simulated assets, the return $r_t$ is not a fixed average of the individuals returns $y_{kt}$ (see Figure~\ref{compo}).

To compare the methods by using formal backtests, we considered the same framework of GARCH estimations on rolling windows of length $1, 000$, but the methods have been backtested on a longer period of length $2, 000$. Moreover, in order to obtain a clearcut comparison between the naive method and the VHS method, the composition  of the portfolio has to be  highly time-varying.  We thus simulated portfolios whose composition alternates as follows: we take an equal proportion of the returns of the even assets $\epsilon_{2k,t}$ during a period of length 100, and then we switch to an equal proportion of the odd assets $\epsilon_{2k+1,t}$ during another period of length 100. Table \ref{Backtests} summarizes the results of the 4 VaR estimation methods for $m=2,4, 8, 100$.
This simulation exercise is intensive since 2000 DCC-GARCH models were estimated for each of the two multivariate methods, and 2000 univariate GARCH(1,1) models were estimated for each of the univariate methods.  The spherical and FHS methods become rapidly too time consuming when the number $m$ of returns increases, because multivariate $m$-GARCH models have to be estimated. Interestingly, the numerical complexity of the univariate methods does not increase much with $m$, so that Table \ref{Backtests} reports  results on portfolios of $m=8$ and $m=100$ assets for the univariate methods only.

In this table, the column Viol gives the relative
frequency of violations (in $\%$), while the columns LRuc, LRind and LRcc give respectively the $p$-values of the unconditional coverage test that the probability of violation is equal to the nominal $5\%$ level, the independence test that
the violations are independent and the conditional coverage test of Christoffersen (2003).
Conclusions drawn from those backtests, which solely focus on the violations, are that all methods are validated on these experiments. 
In particular, it is interesting to notice that the naive method does not behave so poorly in terms of backtests.
It is necessary to introduce alternative statistics to differentiate the different approaches.
From a portfolio's manager point of view, it is interesting to minimize the average VaR (denoted $\overline{\mbox{VaR}}$ in the table) in order to minimize the reserves.
For both regulator and manager, it is also important to minimize the amount of violation. 
The column AV displays the average amount of violation, and the column ES gives the expected shorfall, that is the average loss when the VaR is violated:
for each estimator $\widehat{VaR}_t$ of the conditional VaR, let
$$\mbox{AV}=\frac{\sum_{t=1}^n-(r_t+\widehat{VaR}_t)\mathbf{1}_{\{r_t<-\widehat{VaR}_t\}}}{\sum_{t=1}^n\mathbf{1}_{\{r_t<-\widehat{VaR}_t\}}},
\qquad
\mbox{ES}=\frac{\sum_{t=1}^n-r_t\mathbf{1}_{\{r_t<-\widehat{VaR}_t\}}}{\sum_{t=1}^n\mathbf{1}_{\{r_t<-\widehat{VaR}_t\}}}.
$$
These statistics clearly show that the naive approach is inefficient compared to its competitors.
With this method, the amount of violation tends to be higher whatever the size $m$ of the portfolio. For these statistics AV and ES, the VHS approach appears comparable to the multivariate methods when comparison is possible, that is when $m$ is not too large.
Alternative comparisons are provided by introducing the loss function
$$\mbox{Loss}=\frac1n \sum_{t=1}^n-(r_t+\widehat{VaR}_t)(\alpha-\mathbf{1}_{\{r_t<-\widehat{VaR}_t\}}),$$
also considered by Giacomini and Komunjer (2004), and Gneiting (2011) in the context of forecast evaluation.
The last column of Table \ref{Backtests} reports, for each of the three non-naive methods,
$p$-values of the Diebold-Mariano (1995) test for the null that
the naive method produces the same loss against the alternative that it induces higher loss.
The null is rejected in each situation, leading to the same conclusion as before: the naive method is outperformed by its three competitors when $m$ is small, and by the
VHS method when $m$ is large.

\begin{table}
\caption{\label{Backtests}\footnotesize
Backtests of the 5\%-VaR estimates for the factor model. The
composition of the portfolio changes every 100 observations. For the multivariate (resp. univariate) approaches DCC (resp. GARCH(1,1)) models are estimated
on 1,000 observations. All statistics are
computed on 2,000 observations.
}
	\begin{center}
			\begin{tabular}{ll cc cc ccccc}
\hline\hline
     &  & Viol$^a$ & LRuc$^b$ & LRind$^b$ & LRcc$^b$ & $\overline{\mbox{VaR}}^c$& AV$^d$& ES$^d$& Loss$^d$& DM$^e$ \\
$m=2$ &    Naive     &     5.20  &     68.34 &     79.24 &     88.89 &     4.64  &     1.87  &     6.56  &     0.33  &     -     \\
	 &    VHS    	&     5.55  &     26.71 &     72.63 &  	  50.81 &     4.26  &     1.07  &     5.30  &     0.27  &  5.e-10\\
	 &    Spherical    	&     5.30  &     54.19 &     86.71 &  	  81.87 &     4.28  &     1.10  &     5.49  &     0.27  &  1.e-09\\
	 &    FHS    	&     5.60  &     22.67 &     90.67 &  	  47.82 &     4.25  &     1.10  &     5.33  &     0.27  &  3.e-09\\
$m=4$ &	  Naive     &    5.55  &     26.71 &     12.95 &     17.12 &     4.51  &     1.90   &     6.08  &     0.33  &     -      \\
&    VHS    	&     4.35 & 17.30 &   90.93  &	39.26  & 4.60  &1.18  & 5.45   & 0.28   &  1.e-07\\
&    Spherical    	& 5.20 & 68.34 & 5.95     &15.59 & 4.36  & 1.19 & 5.61  & 0.28 &  2.e-08\\
&    FHS    	& 4.30 & 14.15  &  87.19 &  33.50& 4.60 & 1.19 & 5.55  &  0.28 & 2.e-07\\
$m=8$&    Naive 	  &     5.10   &     83.79 &     56.34 &     82.87 &     4.87  &     1.61  &     6.10   &     0.33  &     -     \\
     &    VHS    &       5.50    &     31.23 &     69.02 &     55.44 &   4.44  &    1.05   &     5.38   &   0.28   & 1.e-08 \\  	
$m=100$& Naive  &     4.90  &     83.69 &     6.93  &     18.8  &     4.53  &     2.16  &     7.65  &     0.34  &     -     \\
    &VHS&            5.25   &     61.07 &     81.42 &    85.46  &     4.56  &     1.09  &     5.61  &  0.29     &   6.e-09 \\
	 	\hline\hline
\multicolumn{11}{l}{\footnotesize{$^a$ \! \% of violations of the estimated 5\%-VaRs; $^b$ \!$p$-values of backtests of the  violations;
$^c$ \!average VaR; }}\\
\multicolumn{11}{l}{\footnotesize{
$^d$ \!average amount of violation criteria; $^e$ \!$p$-values of DM tests that the naive method has the same loss.}}
			\end{tabular}
		\end{center}
\end{table}

\subsection{Real data}
\begin{figure}
\vspace*{10cm} \protect \includegraphics{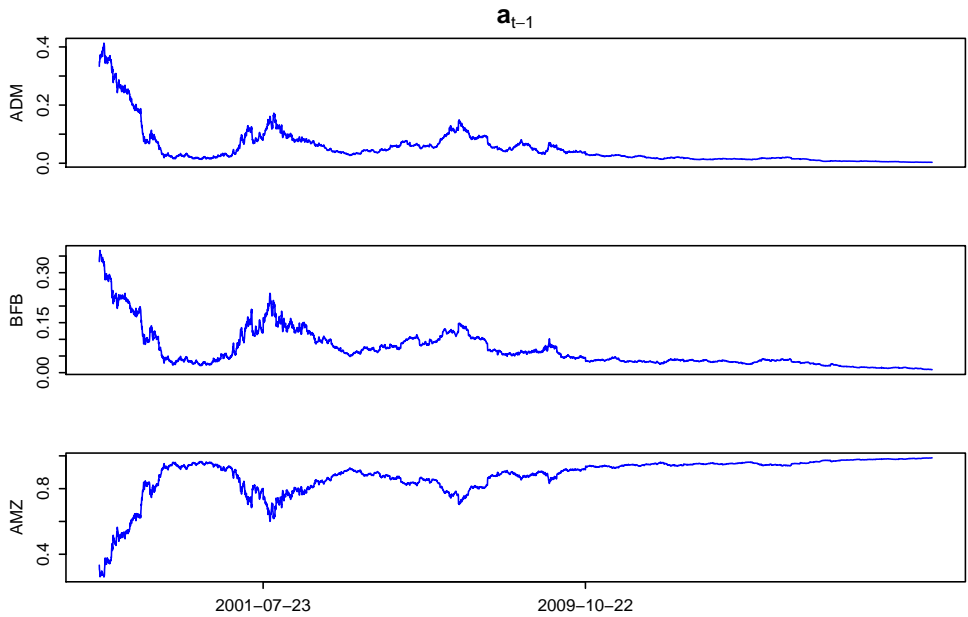}
\vspace{-3cm}
\caption{\label{as.composition1}{\small Time-varying composition of the crystallized portfolio based on the three US stocks.}}
\end{figure}

\begin{figure}
\begin{center}
\vspace*{10cm} \protect \includegraphics{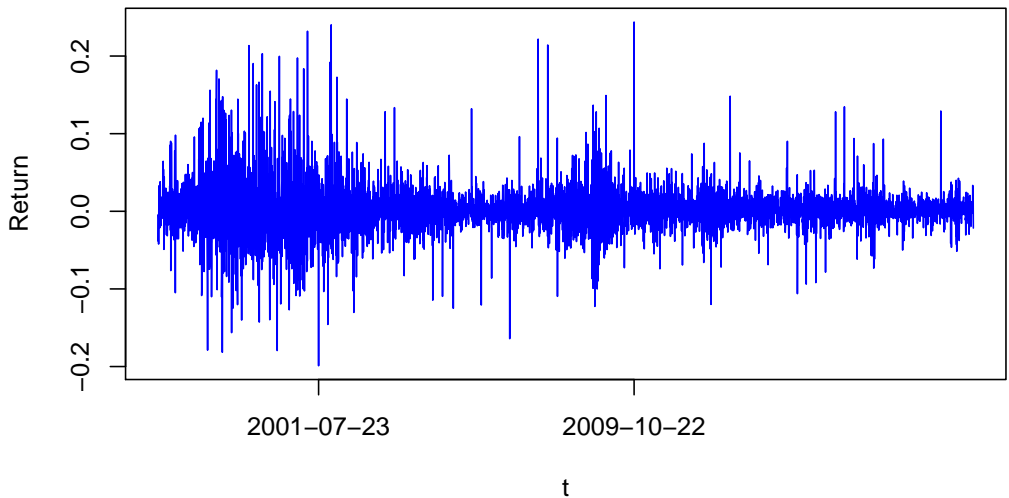}
\end{center}\vspace{-4cm}
\caption{\label{as.epsilon1}{\small Returns of the crystallized portfolio  based on the three US stocks.}}
\end{figure}

We start by plotting the returns of an actual crystallized portfolio, showing the same kind of behaviour as the simulated portfolio of Figure \ref{epsilon1}.
The portfolio  is obtained by equally weighting 3 stocks, ADM (Advanced Micro Devices), BFB (Brown-Forman Corporation) and
AMZ (Amazon).
on the period 1997-05-15 to 2018-09-05 (5363 observations).
The composition $\ba_{t-1}$ of the portfolio is plotted in Figure \ref{as.composition1}.
It is seen that the third asset becomes preponderant at the end of the period.
The plot of the portfolio's returns in Figure \ref{as.epsilon1} shows that the changes in the portfolio composition induce apparent non-stationarities.
Contrary to the simulated portfolio of Figure \ref{epsilon1}, the volatility decreases when the portfolio becomes more concentrated, which is explained by the fact
the third asset is also the less volatile.
Next, we will compare the univariate methods on portfolios whose composition is strongly time varying.
Then, we will consider portfolios that are regularly rebalanced so that their compositions do not vary too much.

\subsubsection{Estimating the conditional VaR of portfolios of US stocks}

\begin{figure}
\vspace*{9cm}
\includegraphics{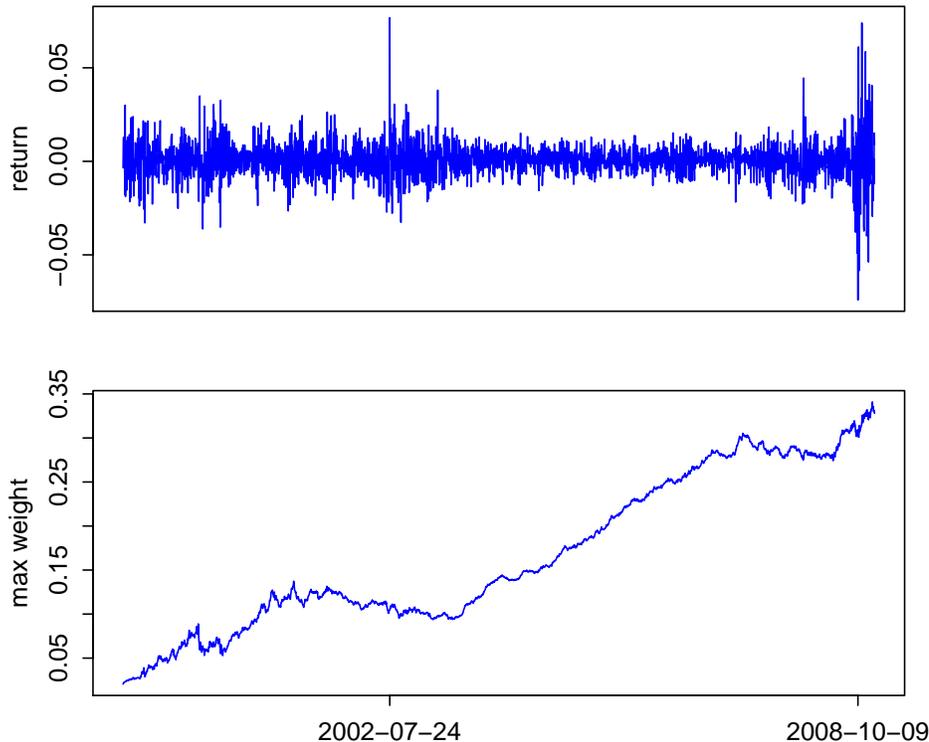}
\caption{\label{CompoPort} {\small
Returns and maximum weight of a crystallized portfolio of 49 US stocks.}}
\end{figure}

We now consider portfolios built from a set of $m=49$ US stocks covering 2,489 trading days, from January 4, 1999 to December 31, 2008.
The data have been kindly provided to us by Sébastien Laurent, and are described in Laurent, Lecourt and Palm  (2016).
The top panel of Figure \ref{CompoPort} displays the returns of a crystallized portfolio which was fully diversified at the beginning of the period,
{\em i.e} with composition $\ba_{0}=(1/m,\dots,1/m)$ at time $t=1$, and for which the number of units of each asset $\mu_{i,t}=1$ is time-invariant. The bottom panel of this figure displays $\max_{i\in\{1,\dots,49\}}a_{i,t}$ as function of $t$. This figure shows that the composition of the portfolio is time-varying, and that the portfolio tends to become more and more concentrated. At the beginning of the period, the return of the crystallized  portfolio is an equi-weighted average of the individual returns, but at the end of the period, one of the individual returns tends to have a prominent weight (this individual return is the UNILEVER stock from  February 20, 2003  onwards).

Given the large number of assets, we did not implement multivariate approaches to estimate the VaR of this portfolio. Estimating a multivariate GARCH(1,1) model by QML in this setting would require inverting very large
correlation/covariance matrices at each step of the optimization algorithm.
There exists multivariate approaches--either based on constrained models or using alternatives to the full QML (e.g. the composite likelihood as in Engle, Ledoit and Wolf (2017), or the Equation-by-Equation method of Francq and Zakoian (2016)), or using intraday data (e.g. Boudt, Laurent, Quaedvlieg, and Sauri (2017))--which do not fit into our semiparametric GARCH framework.

\begin{center}
\begin{figure}
\vspace*{8.5cm}
\includegraphics{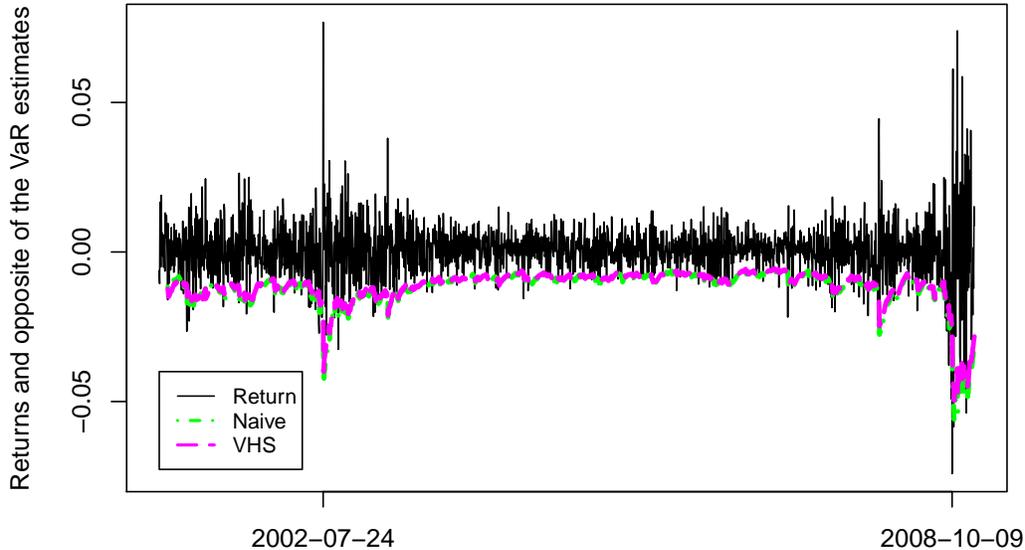}
\caption{\label{2VaRUS} {\small
Naive and VHS estimations of the $5\%$-VaR at horizon 1 for a  portfolio of  US stocks.}}
\end{figure}
\end{center}

Figure~\ref{2VaRUS} displays the estimates of the 5\%-VaR obtained from the Naive and VHS methods. Starting from $t=1, 001$, the estimates are computed from all the previous observations $r_1,\dots,r_{t-1}$.
As can be seen from the figure, the naive and VHS methods provide very similar results and the backtests used in the previous section are not able to distinguish them. At first sight, this result is quite surprising, but it can be explained by the fact that the composition of the portfolio varies relatively slowly and, even if the composition is changing, the dynamics does not change drastically because most of the individual returns follow similar GARCH models. The interesting conclusion is that, even if the naive method is not supported by rigorous theoretical results, it may work surprisingly well in practice.

In a second experiment, we considered a portfolio whose composition changes every week, between  an equi-weighted average of the stocks MS, F, GM  and an  equi-weighted average of the stocks CVX, XOM, ECX. 
At the beginning of the week, the portfolio is thus composed of the same amount of the three assets, then the portfolio is crystallized until the end of the week.
The two sets of stocks have been chosen because of their different dynamic behaviors. Results are displayed in Table~\ref{BacktestsUS}. Once again, the naive  and VHS methods are not much different, but there are some discrepancies in favor of the VHS method (as indicated by the DM test).

\begin{table}\caption{\label{BacktestsUS}\footnotesize
Backtests of estimates of the VaR at the level $\alpha$ obtained by univariate methods, for a  portfolio of  US stocks whose composition changes every week over the period 1999-2008.}
	\begin{center}
			\begin{tabular}{ll cc cc ccccc}
\hline\hline
     &  & Viol & LRuc & LRind & LRcc & $\overline{\mbox{VaR}}$& AV& ES& Loss& DM \\
$\alpha=5\%$ &    Naive     &     6.234  &      \phantom{0}1.484  &      91.914 &      11.426 &      0.024  &      0.012  &      0.036  &      1.9e-03  &      -      \\
	 &    VHS    	&      5.782  &   11.812  & 78.518  &  47.234 &  0.024  &  0.010  &  0.035 & 1.7e-03 & 4e-04\\
$\alpha=1\%$& Naive   &      1.307  &      18.858 &      40.646 &      49.032 &      0.038  &      0.021  &      0.061  &      \phantom{0.}6e-04  &      -     \\
	 &    VHS    	&     1.207  & 36.968 & 29.420  & 59.235 & 0.036&  0.016 & 0.057 & \phantom{0.}5e-04 & 7e-03   \\
	 	\hline\hline
			\end{tabular}
		\end{center}
	\end{table}

\subsubsection{Comparing naive and VHS methods on rebalanced portfolios}
\label{MonteCarlo}
\begin{figure}
\vspace*{10cm} \protect \includegraphics{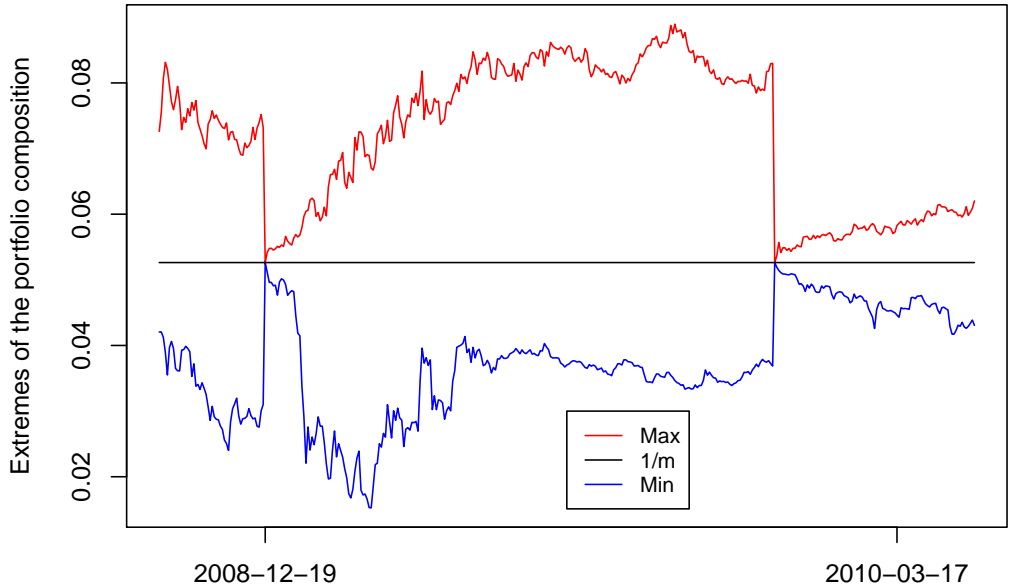}
\vspace{-2cm}
\caption{\label{Extremes.port2}{\small Maximal (in red) and minimal (in blue) value of the composition $\ba_{t-1}$ of a portfolio that
is re-balanced every $T=250$ days.}}
\end{figure}

\begin{figure}
\vspace*{10cm} \protect \includegraphics{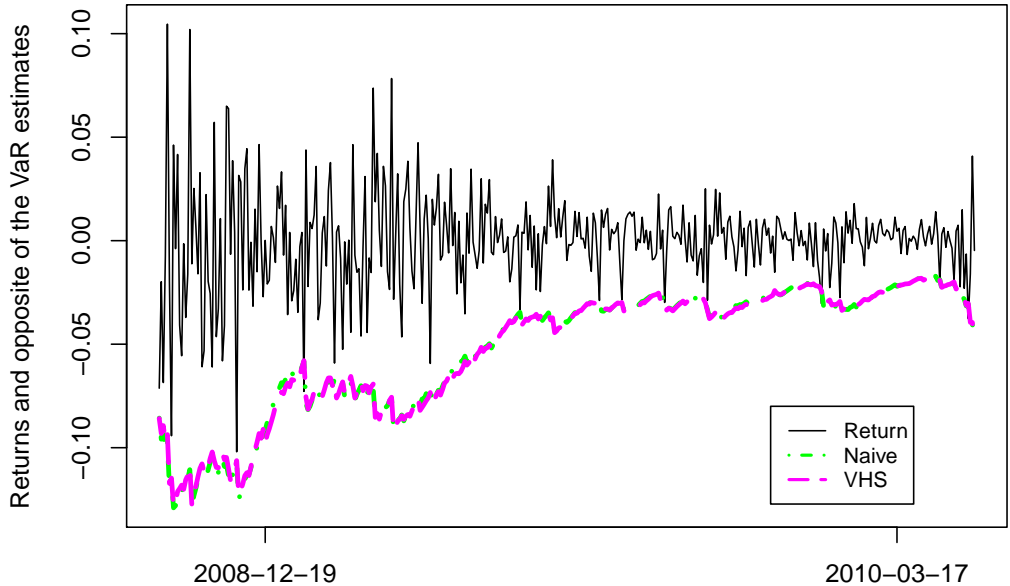}
\vspace{-2cm}
\caption{\label{port2}{\small Estimated $1\%$-VaR by the naive (dotted green line) and VHS (magenta line) methods}}
\end{figure}

In practice, portfolios are often periodically re-balanced. Intuitively, the naive and VHS approaches should behave similarly in this situation. To verify this intuition, we now build portfolios with the $m=19$ stocks that illustrate Chapter 17 of Boyd  and  Vandenberghe (2018). The data set covers the period from 2004-01-02 to 2013-12-31 (2517 values). We consider a portfolio which is equally diversified
at the beginning of the period ({\it i.e.} we take $\mu_{i,t}=\mu_{i}=V_0/(mp_{i,0})$ at $t=0$, corresponding to the beginning date 2004-01-02), and we re-balance the portfolio every $T$ periods (such that $\mu_{i,t}=V_t/(mp_{i,t})$ at $t=kT$ for all $k\in \mathbb{N}$). If the portfolio is re-balanced at any time, {\it i.e.} $T=1$, then $a_{i,t}=1/m$ for all $t$, and thus the naive and VHS methods coincide. To limit transaction costs, the investor can maintain the same asset allocation for an extended period of time. The so-called lazy portfolios or permanent portfolios are re-balanced every year or at any time its asset allocation strays too far from its initial state.
Figure~\ref{Extremes.port2} shows that, when the portfolio is re-balanced every year, the composition of the portfolio can deviate much from the fully diversified portfolio (for which $a_{i,t}=1/m$ for all $i\in\{1,\dots,m\}$).  This does not necessarily entail a huge difference between the VaR's estimated by the naive and VHS methods.
Actually, for the nominal risk level $\alpha=1\%$, the maximum difference between the estimated VaR has been observed on 2008-12-19, with a naive VaR of 0.09163292 and a VHS VaR of 0.09775444. As illustrated by Figure~\ref{port2} the two estimated VaR are hardly distinguishable. This is interesting because it entails that the asymptotic theory built for the VHS method should also apply for the naive method. In other words, the naive method is not so naive if the portfolio is re-balanced from time to time, as recommended by finance professionals.

\section{Conclusion}\label{seccon}
This paper developed a method for estimating the conditional VaR of a portfolio of asset returns, without
relying on a joint dynamic modelling of the vector of returns.
For large portfolios,  using optimization routines in multivariate approaches often entails formidable numerical difficulties.
By circumventing the dimensionality curse, univariate methods provide an operational alternative.

The naive method discussed in this paper 
has no theoretical grounds because it implicitly, and erroneously when the composition is time varying, relies on the stationarity of the returns process.
In many cases, however, it behaves satisfactorily as our numerical experiments revealed.
For the VHS method, we developed an asymptotic theory for a general class of dynamic models, which are not directly estimated on observations but
rather on reconstituted returns. The obtained asymptotic results allow to  quantify the estimation risk
that should be taken into account in risk management.
From our numerical experiments, the multivariate methods can be recommended when the size of the portfolio is small
and the estimated multivariate GARCH model is likely to be well-specified.
When the number of underlying assets is large, or when finding an appropriate multivariate specification is difficult,
the univariate methods offer a valuable alternative. The VHS method requires computing virtual returns, which is negligible computational burden. Thus, the two univariate methods display similar numerical complexity. However, when the naive and VHS estimators differ substantially,
the former should not be considered as reliable. When they provide similar results, the asymptotic results obtained for the VHS method
could  also be used to evaluate the accuracy of the naive method.


\appendix

\begin{center}
\Large{\bf Appendices}
\end{center}

\section{Multivariate approaches}\label{secMul}

We assume in this appendix that the vector of log-returns follows a general multivariate model
of the form 
\begin{equation}\label{garchmulfort}\by_t=\bm_t(\bvartheta_0)+ \bepsilon_t, \qquad \bepsilon_t=\bSigma_t(\bvartheta_0)\beeta_t,\end{equation}
where $(\beeta_t)$ is a sequence of independent and identically
distributed (iid) $\mathbb{R}^m$-valued variables with zero mean and
identity covariance matrix; $\beeta_t$ is independent from the $\by_{t-i}$ for $i>0$;
the $m\times m$ non-singular matrix
$\bSigma_t(\bvartheta_0)$ 
and the $m\times 1$ vector  $\bm_t(\bvartheta_0)$
are 
functions of the past values of $\by_t$ which are parameterized
by a $d$-dimensional parameter $\bvartheta_0$:
\begin{equation}\label{garchmulfort1}
\bm_t(\bvartheta_0)=\bm(\by_{t-1}, \by_{t-2},
\ldots , \bvartheta_0), \qquad \bSigma_t(\bvartheta_0)=\bSigma(\by_{t-1}, \by_{t-2},
\ldots , \bvartheta_0).\end{equation}
Multivariate approaches require specifying the first two conditional moments in (\ref{garchmulfort1})  of the vector of individual returns.
While the conditional mean is generally modelled using a small-order AR process, there are plenty
of GARCH-type specifications for the conditional variance. See for instance
Bauwens, Laurent and
Rombouts (2006), 
Francq and Zakoïan (2010, Chapter 11) or Bauwens, Hafner and Laurent (2012) for presentations of
the most commonly used specifications.

In view of (\ref{garchmulfort}) and (\ref{portfo}), the portfolio's return satisfies
\begin{equation}\label{portfogarch}
r_t=\mathbf{a}'_{t-1}\bm_t(\bvartheta_0)+\mathbf{a}'_{t-1}\bSigma_t(\bvartheta_0)\beeta_t,
\end{equation}
from which it follows that its conditional VaR at level $\alpha$ is given by
\begin{equation}\label{generalvar}
\mbox{VaR}_{t-1}^{(\alpha)}(r_t)= -\mathbf{a}'_{t-1}\bm_t(\bvartheta_0)+\mbox{VaR}_{t-1}^{(\alpha)}\left(\mathbf{a}'_{t-1}\bSigma_t(\bvartheta_0)\beeta_t\right).
 \end{equation}

\subsection{Conditional VaR estimation under conditional ellipticity}
\label{sec3}
The VaR formula can be simplified if we assume that the errors
$\beeta_t$ have a spherical distribution, that is, 
for any non-random vector $\blambda\in
\mathbb{R}^m$, $\blambda'\beeta_t\stackrel{d}{=}
\|\blambda\|\eta_{1t}$,
where $\|\cdot\|$ denotes the euclidian norm on $\mathbb{R}^m$,
$\eta_{it}$ denotes the $i$-th component of $\beeta_t$, and
$\stackrel{d}{=}$ stands for the equality in
distribution.
 Note that assuming sphericity of the distribution of $\beeta_t$ amounts to assuming ellipticity of the conditional distribution of $\by_t$ satisfying (\ref{garchmulfort})-(\ref{garchmulfort1}).
Under the sphericity assumption we have
\begin{equation}\label{sphervar}
\mbox{VaR}_{t-1}^{(\alpha)}(r_t)= -\mathbf{a}'_{t-1}\bm_t(\bvartheta_0)+\left\|\mathbf{a}'_{t-1}\bSigma_t(\bvartheta_0)\right\|\mbox{VaR}^{(\alpha)}\left(\eta\right),
 \end{equation}
where $\mbox{VaR}^{(\alpha)}\left(\eta\right)$ is the (marginal)
VaR of $\eta_{1t}$. Under the sphericity assumption, by \eqref{sphervar} a natural strategy for estimating the conditional VaR
 of a portfolio is to estimate
$\bvartheta_0$ by some consistent estimator $\widehat{\bvartheta}_n$ in a first step, to
extract the residuals
and to estimate $\mbox{VaR}^{(\alpha)}\left(\eta\right)$ in a second step.

An estimator
of the conditional VaR at level $\alpha$ accounting for the conditional ellipticity is thus
\begin{equation}\label{VaRunderS}\widehat{\mbox{VaR}}^{(\alpha)}_{S,t-1}(r_t) = -\mathbf{a}'_{t-1}\widetilde{\bm}_t(\widehat{\bvartheta}_n)+ \|\mathbf{a}'_{t-1}\widetilde{\bSigma}_t(\widehat{\bvartheta}_n)\|\xi_{n,1-2\alpha},
\end{equation}
where $\xi_{n,1-2\alpha}$ is the empirical $(1-2\alpha)$-quantile of all components of the residuals $\widehat{\beeta}_t=\widetilde{\bSigma}_t^{-1}(\widehat{\bvartheta}_n)\{\by_t-\widetilde{\bm}_t(\widehat{\bvartheta}_n)\}$.
 \footnote{By assumption, the components of $\beeta_t$ have the same symmetric distribution.}
 Here $\widetilde{\bm}_t(\widehat{\bvartheta}_n)$ and
$\widetilde{\bSigma}_t(\widehat{\bvartheta}_n)$ denote the estimated conditional mean and variance of $\by_t$ based on initial values
$\widetilde{\by}_i$ for $i\le 0$.
Francq and Zakoïan (2018) derived, under appropriate assumptions, the
asymptotic joint distribution of $\widehat{\bvartheta}_{n}$ and $\xi_{n,1-2\alpha}$.

\subsection{Conditional VaR estimation without the sphericity assumption}
\label{without}

The 
FHS approach (see Barone-Adesi, Giannopoulos and Vosper (1999), Mancini and Trojani (2011) and the references therein) does not require any symmetry assumption.
It relies on estimating the conditional quantile of a linear combination of the components of the innovation, where the coefficients depend on both the model parameter and the past returns.
Indeed, the conditional VaR of the portfolio return is
$$ \mbox{VaR}^{(\alpha)}_{t-1}(r_t)= \mbox{VaR}^{(\alpha)}_{t-1}\left\{\mathbf{a}'_{t-1}\bm_t(\bvartheta_0)+\mathbf{a}'_{t-1}\bSigma_{t}(\bvartheta_0)\beeta_{t}\right\}.
$$
A natural estimator is thus
\begin{equation}
\label{FHS}
\widehat{\mbox{VaR}}^{(\alpha)}_{FHS,t-1}(r_t)
=-q_{\alpha}\left(\left\{\mathbf{a}'_{t-1}\widetilde{\bm}_t(\widehat{\bvartheta}_n)  +\mathbf{a}'_{t-1}\widetilde{\bSigma}_t(\widehat{\bvartheta}_n)\widehat{\beeta}_s, \quad 1\leq s\leq n\right\}\right),
\end{equation}
where $q_{\alpha}(S)$ denotes the $\alpha$-quantile of the elements of any finite set $S\subset \mathbb{R}.$
Francq and Zakoïan (2018)
studied the asymptotic distribution of this estimator.

\section{Proofs}\label{AppendB}


\subsection{Proof of Lemma~\ref{paire}}
We have
$$P\left(\sum_{k=1}^n\Delta_k>c\right)=P(Z_n>c_n)=P(Z_n>0)-P(Z_n\in(0,c_n])\mathbf{1}_{\{c_n>0\}}+P(Z_n\in(c_n,0])\mathbf{1}{c_n\le 0}$$
with $c_n=a_n c+b_n$. We have $P(Z_n>0)\to p$ and,  for any $\varepsilon>0$,
there exists $c_{\varepsilon}>0$ such that
$\lim_{n\to\infty}P(Z_n\in(0,c_n])\leq \lim_{n\to\infty}P(Z_n\in[-c_{\varepsilon},c_{\varepsilon}])\leq \varepsilon$.
The conclusion follows.
\zak

\subsection{Proof of Lemma \ref{exempleGARCH}.}
We start by showing that Assumption {\bf A2} is satisfied.
Let $a(z)=\alpha_0 z^2+\beta_0$ and let
$$\epsilon_t=\sqrt{\omega_0}\left\{1+ \sum_{i=1}^{\infty} a(u_{t-1})\ldots a(u_{t-i})\right\}^{1/2}u_t,$$
which is well defined under the condition in i). The process $(\epsilon_t)$ is strictly stationary and ergodic by {\bf A1}.
The second part of {\bf A2} follows from iv).
Next, we have 
$$\sigma_t^2(\btheta_0)=\sigma_{t,T_n}^2+\widetilde{\sigma}^2_{t,T_n},\qquad \sigma_{t,T_n}^2=\omega_0\left\{1+\sum_{k=1}^{T_n}\prod_{i=1}^{k}a(u_{t-i})\right\}.$$
Note that, under the strict stationarity condition $E\log a(u_1)<0$,  we have
\begin{equation}
\label{beauté}
\widetilde{\sigma}^2_{t,T_n}=\omega_0\sum_{k=T_n+1}^{\infty}\prod_{i=1}^{k}a(u_{t-i})\to 0\mbox{ a.s. when }T_n\to\infty.
\end{equation}
We also set $\epsilon_t^2=\epsilon_{t,T_n}^2+\widetilde{\epsilon}_{t,T_n}^2$, where $\epsilon_{t,T_n}=u_t\sigma_{t,T_n}\in {\cal F}_{t:t-T_n}$. We thus have
$$\sigma_{t,T_n}^2=\sum_{i=0}^{T_n-2}\beta_0^i\left(\omega_0+\alpha_0\epsilon^2_{t-i-1,T_n-i-1}\right)+\beta_0^{T_n-1}\sigma_{t-T_n+1,1}^2.$$
Now, we show that {\bf A3} holds true for the GARCH(1,1) model.
The first and second components of $\bD_t$ are bounded, and thus can be handled easily. The last component of $\bD_t$ has the form
$${_\beta}{\sigma}_{t}^2:=\frac{1}{2\sigma_t^2}\frac{\partial \sigma^2_t(\btheta_0)}{\partial \beta}={_\beta}{\sigma_{t,T_n}^2}+{_\beta}{\widetilde{\sigma}_{t,T_n}^2},\qquad {_\beta}{\sigma_{t,T_n}^2}=\frac{\sum_{i=1}^{T_n-2}i\beta_0^{i-1}\left(\omega_0+\alpha_0\epsilon^2_{t-i,T_n-i}\right)}{\sigma_{t,T_n}^2}.$$
Note that ${_\beta}{\sigma_{t,T_n}^2}\in {\cal F}_{t-1:t-T_n}$ and, using the inequality $x/(1+x)\leq x^s$ for any $x\geq 0$ and any $s\in (0,1)$, we have
$${_\beta}{\sigma_{t,T_n}^2}\leq \frac{1}{(1-\beta_0)^2}+ \sum_{i=1}^{T_n-2}\frac{i\beta_0^{i-1}\alpha_0\epsilon^2_{t-i,T_n-i}}{\omega_0+\beta_0^{i}\alpha_0\epsilon^2_{t-i,T_n-i}}\leq \frac{1}{(1-\beta_0)^2}+ \frac{\alpha_0^s}{\beta_0\omega_0^{s}}\sum_{i=1}^{T_n-2}i\left\{\beta_0^{ s}\right\}^i\epsilon^{2s}_{t-i,T_n-i}.$$
Therefore, under {\bf A2}, for any $r\geq 1$, choosing  $s>0$ such that $2sr\leq s_0$, 
the Hölder inequality shows that
$$\sup_n\left\|{_\beta}{\sigma_{t,T_n}^2}\right\|_r\leq \frac{1}{(1-\beta_0)^2}+ \frac{\alpha_0^s}{\beta_0\omega_0^{s}}\left\|\epsilon^{2s}_{1}\right\|_r\sum_{i=1}^{\infty}i\left\{\beta_0^{ s}\right\}^i<\infty,\quad \left\|{_\beta}{\sigma_{t}^2}\right\|_r<\infty,$$
where $\|\cdot \|_r$ denotes the $L_r$ norm.
 Now note that
\begin{eqnarray}
\label{tousegaux}{_\beta}{\widetilde{\sigma}_{t,T_n}^2}&=&
\frac1{2\sigma_{t,T_n}^2}\frac{\partial \sigma^2_t(\btheta_0)}{\partial \beta}+\frac12\frac{\partial \sigma^2_t(\btheta_0)}{\partial \beta}\left(\frac1{\sigma^2_t(\btheta_0)}-\frac1{\sigma^2_{t,T_n}}\right)
-{_\beta}{\sigma_{t,T_n}^2}
\nonumber\\&=&
\frac1{2\sigma_{t,T_n}^2}\sum_{i=1}^{T_n-2}i\beta_0^{i-1}\left(\omega_0+\alpha_0\epsilon^2_{t-i,T_n-i}\right)+
\frac1{2\sigma_{t,T_n}^2}\sum_{i=1}^{T_n-2}i\beta_0^{i-1}\alpha_0(\epsilon^2_{t-i}-\epsilon^2_{t-i,T_n-i})-{_\beta}{\sigma_{t,T_n}^2}
\nonumber\\
&&+\frac{1}{2\sigma_{t,T_n}^2}\sum_{i=T_n-1}^{\infty}i\beta_0^{i-1}
\left(\omega_0+\alpha_0\epsilon^2_{t-i}\right)+{_\beta}{\sigma_{t}^2}\left(1-\frac{\sigma^2_t(\btheta_0)}{\sigma^2_{t,T_n}}\right)
\nonumber\\
&=&
-\frac{\widetilde{\sigma}_{t,T_n}^2}{\sigma_{t,T_n}^2}{_\beta}{\sigma_{t}^2}+
\frac{\alpha_0\sum_{i=1}^{T_n-2}i\beta_0^{i-1}\widetilde{\epsilon}^2_{t-i,T_n-i}}{2\sigma_{t,T_n}^2}+
\frac{\sum_{i=T_n-1}^{\infty}i\beta_0^{i-1}
\left(\omega_0+\alpha_0\epsilon^2_{t-i}\right)}{2\sigma_{t,T_n}^2}.
\end{eqnarray}
In view of \eqref{beauté}, the first term of the right-hand side of the equality tends to zero in probability.
Using Lemma 2.2 in 
Francq and Zakoïan (2010),
the strict stationarity condition $E\log a(u_1)<0$ entails the existence of $s\in (0,1)$ such that $\rho:=Ea^s(u_1)<1$. We then have $E\widetilde{\epsilon}_{t,T_n}^{2s}\leq K\rho^{T_n}$, which entails
$$E\left|\sum_{i=1}^{T_n-2}i\beta_0^{i-1}\widetilde{\epsilon}^2_{t-i,T_n-i}\right|^s
\leq K\sum_{i=1}^{T_n-2}i^s\beta_0^{s(i-1)}\rho^{T_n-i}\to 0\quad \mbox{ as }\quad n\to\infty.$$
Noting that $E|X_n|^s\to 0$ for some $s>0$ entails that $X_n\to 0$ in probability, we conclude that the second term of the right-hand side of the equality \eqref{tousegaux} tends to zero in probability.
Let $s\in(0,1)$ such that $E|\epsilon_t|^{2s}<\infty$. We have
$$E\left|\sum_{i=T_n-1}^{\infty}i\beta_0^{i-1}
\left(\omega_0+\alpha_0\epsilon^2_{t-i}\right)\right|^s\leq \left(\omega_0^s+\alpha^s_0E|\epsilon_{1}|^{2s}\right) \sum_{i=T_n-1}^{\infty}i^s(\beta_0^s)^{i-1}
\to 0$$ as $n\to\infty$.
If follows that the third term of the right-hand side of the equality \eqref{tousegaux} tends to zero in probability.
We thus have shown that ${_\beta}{\widetilde{\sigma}_{t,T_n}^2}=o_p(1).$
It follows that
${_\beta}{\sigma_{t,T_n}^2}$ can be chosen as being the last component of $\bD_{t,T_n}$. As already argued, the two other components are handled more easily.
This shows that {\bf A3} is satisfied for any sequence $T_n$ tending to infinity.
Turning to {\bf A4}, assume that $\sigma_{t+1}(\btheta_1)=\sigma_{t+1}(\btheta_2)\; a.s.$ for $\btheta_i=(\omega_i, \alpha_i, \beta_i)$. We thus have
$\omega_1+\alpha_1u_{t}^2\sigma_{t}^2(\btheta_0)+ \beta_1\sigma_{t}^2(\btheta_1)=
\omega_2+\alpha_2u_{t}^2\sigma_{t}^2(\btheta_0)+ \beta_2\sigma_{t}^2(\btheta_2)\; a.s.$ Thus, if $\alpha_1\ne \alpha_2$,  $u_t^2$ can be written as a variable belonging to ${\cal F}_{t-1}$. This variable is in fact degenerate and equal to 1
because $E(u^2_t\mid {\cal F}_{t-1})=1$, in contradiction with ii). It follows that $\alpha_1= \alpha_2$, and thus
$\omega_1+ \beta_1\sigma_{t}^2(\btheta_1)=
\omega_2+ \beta_2\sigma_{t}^2(\btheta_2)$. Proceeding in the same way, by expressing $u_{t-1}^2$ as a ${\cal F}_{t-2}$-measurable variable, allows to conclude that $\btheta_1=\btheta_2$, that is that {\bf A4} is satisfied.
The other assumptions are easily verified.
In particular,  {\bf A7} can be handled using (7.51) and (7.54) in Francq and Zakoian (2010)\footnote{For the latter equation, examination of the proof shows  that the iidness of the innovation is not used.}.
\zak

\subsection{Proof of Theorem~\ref{hard}}

We start by showing the following lemma, which only requires a small part of the assumptions of Theorem \ref{hard}.  
\begin{lem}\label{consist}
Assume that 
$E(u^2_t\mid {\cal F}_{t-1})=1$, and that Assumptions
{\bf A2, A4} and {\bf A5} hold. Then $\hat{\btheta}_n\to\btheta_0$ a.s. as $n\to \infty$.
\end{lem}
\noindent
{\bf Proof.} Let
$$Q_n(\theta)=
\frac1n \sum_{t=1}^n {\ell}_t,
\qquad \ell_t={\ell}_t(\btheta)= \frac{\epsilon_t^2}{{\sigma}_t^2(\btheta)}+\log {\sigma}_t^2(\btheta).$$
The strong consistency of $\hat{\btheta}_n$
 is a consequence of the following  results:
\begin{eqnarray*}
&&i) \; \lim_{n\to \infty} \sup_{\theta\in\Theta}|
Q_n(\theta)-\tilde{Q}_n(\theta)|=0\;, \quad a.s.
\\
&& ii) \; \mathbb{E}|\ell_t(\theta_0)|<\infty,  \mbox{ and if } \theta\ne \theta_0\;, \quad
\mathbb{E}\ell_t(\theta)> \mathbb{E}\ell_t(\theta_0)\;;\\
&& iii) \; \mbox{any $\theta\neq \theta_0$ has a neighborhood
$V(\theta)$ such that } \quad 
\limsup_{n\to\infty}\inf_{\theta^*\in
V(\theta)}{Q}_n(\theta^*)> \mathbb{E}\ell_t(\theta_0)
\;, \;\; a.s.
\end{eqnarray*}
To prove i) we note that
\begin{eqnarray*}
 \sup_{\theta\in\Theta}|
Q_n(\theta)-\tilde{Q}_n(\theta)|&\le& n^{-1} \sum_{t=1}^n  \sup_{\theta\in\Theta}\left\{ \left|\frac{\sigma_t^2(\btheta)-\widetilde{\sigma}_t^2(\btheta)}{\sigma_t^2(\btheta)\widetilde{\sigma}_t^2(\btheta)}\right|\epsilon_t^2
+\left|\log \left(\frac{\sigma_t^2(\btheta)}{\widetilde{\sigma}_t^2(\btheta)}\right)\right|
\right\}\\
&\le & \frac{2C}{\underline{\omega}^3} n^{-1} \sum_{t=1}^n \rho^t \epsilon_t^2+\frac{2C}{\underline{\omega}} n^{-1} \sum_{t=1}^n \rho^t,
\end{eqnarray*}
where the latter inequality is deduced from {\bf A4-A5}. The right-hand side
  goes to 0 a.s., by the Ces\`{a}ro lemma and the existence of a small-order moment for $\epsilon_t$ (Assumption {\bf A2}).
  The proof of ii) uses the identifiability assumption in {\bf A4}, and the fact that $E(u^2_t\mid {\cal F}_{t-1})=1$, enabling us to
  write $$\mathbb{E}{\ell}_t(\btheta)= \mathbb{E}\frac{{\sigma}_t^2(\btheta_0)}{{\sigma}_t^2(\btheta)}+\mathbb{E}\log {\sigma}_t^2(\btheta),$$
  where the existence of the latter expectation, in $\mathbb{R}\cup \{+\infty\}$, follows from the first part of {\bf A4}. At the true value we have
  $\mathbb{E}|{\ell}_t(\btheta_0)|<\infty$ using again $E|\epsilon_1|^{s_0}<\infty$.
  The proof of iii) uses the ergodic theorem and a standard compactness argument (see for instance Francq and Zakoian (Proof of Theorem 7.1, 2010) for details).
\zak
Turning to the asymptotic distribution, we establish the following lemma.
\begin{lem}\label{notsimple}
Under {\bf A1} and {\bf A3-A4}, we have
$$\frac{1}{\sqrt{n}}\sum_{t=1}^n\left(\begin{array}{c}(u_t^2-1)\bD_t\\\mathbf{1}_{\{u_t<\xi_{\alpha}\}}-\alpha\end{array}\right)\stackrel{{\cal L}}{\to}{\cal N}(0,\bS_{\alpha}),
\quad \mbox{where} \quad \bS_{\alpha}=\left(\begin{array}{cc}\bS^{11}&\bS_{\alpha}^{12}\\
\bS_{\alpha}^{21}&\bS_{\alpha}^{22}\end{array}\right)$$
is a positive definite matrix.
\end{lem}
\noindent
{\bf Proof.} Let $c_0\in \mathbb{R}$, $\bc_1\in \mathbb{R}^m$, $\bc=(\bc_1', c_0)'$ and
$$x_{t}=(u_t^2-1)\bc_1'\bD_{t}+c_0(\mathbf{1}_{\{u_t<\xi_{\alpha}\}}-\alpha),\quad x_{t,n}=(u_t^2-1)\bc_1'\bD_{t,T_n}+c_0(\mathbf{1}_{\{u_t<\xi_{\alpha}\}}-\alpha).$$
We will apply a central limit theorem 
for the mixing triangular array $(x_{t,n})$.
 For convenience, we reproduce it below.
 \begin{theo}[Francq and Zakoïan (2005)]
 Let $(x_{t,n})$ be a triangular array of centered real-valued random variables. For each $n\ge 2$ and $h=1, \ldots, n-1$, let the strong mixing coefficients of $x_{1, n}, \ldots, x_{n,n}$ be defined by
 $$\alpha_n(h)=\sup_{1\le t\le n-h} \sup_{A\in {\cal A}_{t,n}, B\in {\cal B}_{t+h,n}} |P(A\cap B)-P(A)P(B)|,$$
 where ${\cal A}_{t,n}=\sigma(x_{u,n}, 1\le u\le t)$ and ${\cal B}_{t,n}=\sigma(x_{u,n}, t\le u\le n)$ and, by convention, $\alpha_n(h)=1/4$ for $h\le 0$, $\alpha_n(h)=0$ for $h\ge n$.
 Let $S_n=\sum_{t=1}^n x_{t,n}$. Under the following assumptions
 \begin{enumerate}[label=\arabic*)]
 \item $\sup_{n\ge 1}\sup_{1\le t\le n}\|x_{t,n}\|_{2+\nu^*}<\infty$ for some $\nu^*\in (0, \infty]$,
 \item $\lim_{n\to\infty}n^{-1}\mbox{Var}S_n=\sigma^2>0$,
 \item there exists a sequence of integers $(T_n)$ such that $T_n=O(n^{\kappa})$ for some $\kappa\in [0, \nu^*/\{4(1+\nu^*)\})$ and a sequence $\{\alpha(h)\}_{h\ge 1}$ such that
 \begin{eqnarray}\label{e45}\alpha_n(h)\le \alpha(h-T_n), \quad \mbox{for all} \;\; h>T_n,\end{eqnarray}
 \begin{eqnarray}\label{e46}\sum_{h=1}^{\infty}h^{r^*}\alpha^{\nu^*/(2+\nu^*)}(h)<\infty\mbox{ for some }r^*>\frac{2\kappa(1+\nu^*)}{\nu^*-2\kappa(1+\nu^*)},\end{eqnarray}
 \end{enumerate}
 we have $n^{-1/2}S_n\stackrel{{\cal L}}{\to}{\cal N}(0,\sigma^2).$
 \end{theo}

Let $\|\cdot\|$ denote any norm on $\mathbb{R}^d$ and, for any random vector $\bX\in \mathbb{R}^d$,
let $\|\bX\|_p= E(\|\bX\|^p)^{1/p}$ for $p\ge 1$.
Let $\nu^*=(\nu-\epsilon)/2$ where $\nu$ and $\epsilon$ are defined in {\bf A1}.
By {\bf A1} and {\bf A3}, we have $$\sup_n\|u_t^2\bD_{t,T_n}\|^{2+\nu^*}_{2+\nu^*}\leq \|u_t^{4+2\nu^*}\|_{p}\left\|\sup_n\|\bD_{t,T_n}\|^{2+\nu^*}\right\|_{q}<\infty$$
provided that $p={q}/(q-1)$ satisfies $1< p\le (4+\nu)/(4+\nu-\epsilon).$ 
Therefore 1) holds. Now, letting $\widetilde{\bD}_{t,T_n}=\bD_t-{\bD}_{t,T_n}$, we note that
$\|\widetilde{\bD}_{t,T_n}\|^r\to 0$ in probability by {\bf A3}, and the sequence $\|\widetilde{\bD}_{t,T_n}\|^r$ is uniformly integrable
because
$$\sup_n\|\widetilde{\bD}_{t,T_n}\|_r\leq\sup_n\|{\bD}_{t,T_n}\|_r+\|{\bD}_{t}\|_r<\infty. $$ From Theorem 3.5 in Billingsley (1999) it follows that $E\|\widetilde{\bD}_{t,T_n}\|^r\to 0$ as $n\to\infty$, for any $r\geq 1$.
We thus have
\begin{eqnarray}
\nonumber
\mbox{Var}\frac{1}{\sqrt{n}}\sum_{t=1}^n \left(x_t-x_{t,n}\right)&=&E\left[(u_t^2-1)^2\{\bc_1'\widetilde{\bD}_{t,T_n}\}^2\right]\\
&\leq&
E\left[(u_t^2-1)^{2(1+\frac{\nu}4)}\right]^{\frac4{4+\nu}}E\left[(\bc_1'\widetilde{\bD}_{t,T_n})^{2(1+\frac4{\nu})}\right]^{\frac{\nu}{4+\nu}}
\to 0
\label{grosse}
\end{eqnarray} as $n\to\infty$.
Therefore, for $\bc\ne 0$,
$$\lim_{n\to\infty}n^{-1}\mbox{Var}\sum_{t=1}^n x_{t,n}=\lim_{n\to\infty}n^{-1}\mbox{Var}\sum_{t=1}^n x_{t}\to \sigma^2=\bc'\bS_{\alpha} \bc>0$$
provided that $\bS_{\alpha}$ is positive definite. To show the latter, let $x_0\in \mathbb{R}$, $\bx_1\in \mathbb{R}^m$, $\bx=(\bx_1', x_0)'$
such that $\bx'\bS_{\alpha} \bx=0.$ We then have
$(u_t^2-1)\bx_1'\bD_t+ x_0(\mathbf{1}_{\{u_t<\xi_{\alpha}\}}-\alpha)=0$ a.s. Thus, conditional on ${\cal F}_{t-1}$, $u_t$ takes at most three different values when $\bx\ne 0$, in contradiction with the existence
of the density $f_{t-1}$.   Thus 2) is satisfied.
By {\bf A3}, $x_{t,n}\in {\cal F}_{t:t-T_n}$. Thus, $\bS_{\alpha}$ is positive definite and (\ref{e45}) is satisfied, where the sequence $\alpha(h)$ is defined in {\bf A1}.
The conditions  of 3) are also satisfied by {\bf A1}. 
It follows that
$$\frac{1}{\sqrt{n}}\sum_{t=1}^n x_{t,n}\stackrel{{\cal L}}{\to}{\cal N}(0,\sigma^2).$$
The conclusion then follows from \eqref{grosse} and the Cramér-Wold device.
\zak

Now we turn to the proof of Theorem \ref{hard}.
Let $u_t(\btheta)=\epsilon_t/\sigma_t(\btheta)$ and
$\hat{u}_t=\epsilon_t/\tilde{\sigma}_t(\hat{\btheta}_n)$. Note that, by {\bf A4} and {\bf A5}, for $n$ large enough
\begin{equation}
\label{diffhatet}
\left|\hat{u}_t-u_t(\hat{\btheta}_n)\right|=\left|\epsilon_t\frac{\sigma_t(\hat{\btheta}_n)-\widetilde{\sigma}_t(\hat{\btheta}_n)}
{\widetilde{\sigma}_t(\hat{\btheta}_n)\sigma_t(\hat{\btheta}_n)}\right|\leq \frac{C}{\underline{\omega}}\rho^tu_t\sup_{\btheta \in V(\btheta_0)} \left|\frac{\sigma_t(\btheta_0)}{\sigma_t(\btheta)}\right|.
\end{equation}
A Taylor expansion around $\btheta_0$ and {\bf A4}, {\bf A5} yield
$$\hat{u}_t=u_t-u_t\bD_t'(\hat{\btheta}_n-\btheta_0)+ r_{n,t}
$$
with
$$r_{n,t}=\frac{1}{2}(\hat{\btheta}_n-\btheta_0)'\frac{\partial^2
u_t(\btheta_{n,t})}{\partial
\btheta\partial
\btheta'}(\hat{\btheta}_n-\btheta_0)+\hat{u}_t-u_t(\hat{\btheta}_n),$$
where
$\btheta_{n,t}$ is between $\hat{\btheta}_n$ and $\btheta_0$.
Following the approach of Knight (1998) and Koenker and Xiao (2006) (see also Francq and Zakoïan (2015)), we then obtain
\begin{eqnarray*}\sqrt{n}(\xi_{n, \alpha}-\xi_{\alpha})=\arg\min_{z\in
\mathbb{R}} {\cal Q}_n(z), \quad \mbox{where }\;\;
{\cal Q}_n(z)
=zX_n+I_n(z)+J_n(z)+K_n(z),
\end{eqnarray*}
with
\begin{eqnarray*}
X_n&=&\frac{1}{\sqrt{n}}\sum_{t=1}^n
(\mathbf{1}_{\{u_t<\xi_{\alpha}\}}-\alpha),\\
I_n(z)&=&\sum_{t=1}^n\int_0^{z/\sqrt{n}}(\mathbf{1}_{\{u_t\leq \xi_{\alpha}+s\}}-\mathbf{1}_{\{u_t<\xi_{\alpha}\}})ds,\\
J_n(z)&=&\sum_{t=1}^n \int_{0}^{R_{t,n}/\sqrt{n}}\left(\mathbf{1}_{\{u_t-\xi_{\alpha}-z/\sqrt{n}\leq
u\}}-\mathbf{1}_{\{u_t-\xi_{\alpha}-z/\sqrt{n}<0\}}\right)du,\\
K_{n}(z)&=& \sum_{t=1}^n \frac{R_{t,n}}{\sqrt{n}}\mathbf{1}_{\{u_t-\xi_{\alpha}\in (0,z/\sqrt{n})\}}^{*},
\end{eqnarray*}
$\mathbf{1}_{\{X\in (a,b)\}}^{*}=\mathbf{1}_{\{X<b\}}-\mathbf{1}_{\{X<a\}}$ for any real numbers
$a, b$ and any real random variable $X$,
and
$R_{t,n}=u_tD'_t\sqrt{n}(\hat{\btheta}_n-\btheta_0)-\sqrt{n}r_{n,t}.$
We will show that
\begin{equation}\label{yes}
{\cal Q}_n(z)=\frac{z^2}2Ef_0(\xi_{\alpha})+z\{X_n+\xi_{\alpha}\bPsi_{\alpha}'\sqrt{n}(\hat{\btheta}_n-\btheta_0)\}+O_P(1).
\end{equation}
Noting that
\begin{eqnarray*}\nonumber K_{n}(z)&=&
\left(\frac1{\sqrt{n}}\sum_{t=1}^nu_t\mathbf{1}_{\{u_t-\xi_{\alpha}\in (0,z/\sqrt{n})\}}^{*}D'_t\right)\sqrt{n}(\hat{\btheta}_n-\btheta_0)\\
\nonumber
&&-\sqrt{n}(\hat{\btheta}_n-\btheta_0)'\frac{1}{2n}\sum_{t=1}^n\frac{\partial^2
u_t(\btheta_{n,t})}{\partial
\btheta\partial
\btheta'}\mathbf{1}^*_{\{u_t-\xi_\alpha\in (0,z/\sqrt{n})\}}\sqrt{n}(\hat{\btheta}_n-\btheta_0)\\
&&-\sum_{t=1}^n\left\{\hat{u}_t-u_t(\hat{\btheta}_n)\right\}\mathbf{1}^*_{\{u_t-\xi_\alpha\in (0,z/\sqrt{n})\}}\nonumber\\&=:&
K_{n1}(z)+K_{n2}(z)+K_{n3}(z),
\label{decadis}
\end{eqnarray*}
the proof of \eqref{yes} will be divided in the following steps.
\begin{enumerate}[label=\roman*)]
\item $K_{ni}(z)\to 0$ in probability as $n\to\infty$, for $i=2,3$.
\item $K_{n1}(z)=
z \xi_{\alpha}\bPsi_{\alpha}'\sqrt{n}(\hat{\btheta}_n-\btheta_0)+o_P(1)$
 in probability as $n\to\infty$.
\item $J_n(z)$ does not depend on $z$ asymptotically.
\item $I_n(z)\to\frac{z^2}2Ef_0(\xi_{\alpha})$ in probability
as $n\to\infty$.
\end{enumerate}
To prove  i) for $i=2$, note that
\begin{eqnarray}\label{neweq1}\frac{\partial^2
u_t(\btheta)}{\partial
\btheta\partial
\btheta'}=-u_t\frac{\sigma_t(\btheta_0)}{\sigma_t(\btheta)}\frac{1}{\sigma_t(\btheta)}\frac{\partial^2\sigma_t(\btheta)}{\partial
\btheta\partial
\btheta'}+2u_t\frac{\sigma_t(\btheta_0)}{\sigma_t(\btheta)}\frac{1}{\sigma^2_t(\btheta)}\frac{\partial\sigma_t(\btheta)}{\partial
\btheta}\frac{\partial\sigma_t(\btheta)}{\partial
\btheta'}.
\end{eqnarray}
Let $0<\delta<\frac{2+\nu}{6+\nu}$.
By the Cauchy-Schwartz inequality we have, for $p,q,r>0$ such that $\frac1p+\frac1q+\frac1r=1$,
\begin{eqnarray}
&&E\sup_{\btheta\in V(\btheta_0)}\left\|u_t\frac{\sigma_t(\btheta_0)}{\sigma_t(\btheta)}\frac{1}{\sigma_t(\btheta)}\frac{\partial^2\sigma_t(\btheta)}{\partial
\btheta\partial
\btheta'}\right\|^{1+\delta}\nonumber \\&\le &\{E|u_t|^{p(1+\delta)}\}^{1/p}
\left(E\sup_{\btheta\in V(\btheta_0)}\left|\frac{\sigma_t(\btheta_0)}{\sigma_t(\btheta)}\right|^{q(1+\delta)}\right)^{1/q}
\left(E\sup_{\btheta\in V(\btheta_0)}\left\|\frac{1}{\sigma_t(\btheta)}\frac{\partial^2\sigma_t(\btheta)}{\partial
\btheta\partial
\btheta'}\right\|^{r(1+\delta)}\right)^{1/r}.\label{neweq}
\end{eqnarray}
In view of {\bf A1} and  {\bf A7}, the first and third expectation in the right-hand side are finite if we choose $p=\frac{4+\nu}{1+\delta}$ and $r=\frac{2}{1+\delta}$.
We then have $q(1+\delta)=\frac{2(4+\nu)(1+\delta)}{2+\nu-\delta(6+\nu)}$. The latter term increases, when
$\delta$ varies in $ (0, \frac{2+\nu}{6+\nu})$, from $\frac{2(4+\nu)}{2+\nu}$ to infinity.
It is thus possible to choose $\delta$ small enough such that $q(1+\delta)<\frac{2(4+\nu)(1+\tau)}{2+\nu}$.
For such $\delta$, by {\bf A7}, the second expectation and finally the right-hand side of (\ref{neweq}) are finite.
The second summand in the right-hand side of (\ref{neweq1}) can be handled similarly. Thus we have
\begin{equation}
\label{momentnu}E\sup_{\btheta\in V(\btheta_0)}\left\|\frac{\partial^2
u_t(\btheta)}{\partial
\btheta\partial
\btheta'}\right\|^{1+\delta}<\infty.
\end{equation}
By the Hölder inequality, for $\btheta_{n,t}\in V(\btheta_0)$,
\begin{eqnarray*}
&&\left\|\frac{1}{n}\sum_{t=1}^n\frac{\partial^2
u_t(\btheta_{n,t})}{\partial
\btheta\partial
\btheta'}\mathbf{1}^*_{\{u_t-\xi_\alpha\in (0,z/\sqrt{n})\}}\right\|\\
&\leq&\left\{\frac{1}{n}\sum_{t=1}^n\sup_{\btheta\in V(\btheta_0)}\left\|\frac{\partial^2
u_t(\btheta)}{\partial
\btheta\partial
\btheta'}\right\|^{1+\delta}\right\}^{1/(1+\delta)}\left\{\frac{1}{n}\sum_{t=1}^n\mathbf{1}^*_{\{u_t-\xi_\alpha\in (0,z/\sqrt{n})\}}\right\}^{\delta/(1+\delta)}.
\end{eqnarray*}
By \eqref{momentnu} and the ergodic theorem, the limit of the first term of the latter product is almost surely finite.
Letting $\nu_{t,n}=\mathbf{1}^*_{\{u_t-\xi_\alpha\in (0,z/\sqrt{n})\}}$ and $\overline{\nu}_{t,n}=\nu_{t,n}-E(\nu_{t,n}\mid {\cal F}_{t-1})$, we have
$$\frac{1}{\sqrt{n}}\sum_{t=1}^n\mathbf{1}^*_{\{u_t-\xi_\alpha\in (0,z/\sqrt{n})\}}=\frac{1}{\sqrt{n}}\sum_{t=1}^n\overline{\nu}_{t,n}+\frac{1}{\sqrt{n}}\sum_{t=1}^nE(\nu_{t,n}\mid {\cal F}_{t-1}).$$
First note that
$$E(\nu_{t,n}\mid {\cal F}_{t-1})=
\int_{\xi_{\alpha}}^{\xi_{\alpha}+z/\sqrt{n}}f_{t-1}(x)dx=\frac{z}{\sqrt{n}}f_{t-1}(\xi_{\alpha})+\frac{k_{t,n}}{\sqrt{n}}
$$
where
$$\left|k_{t,n}\right|=\sqrt{n}\left|\int_{\xi_{\alpha}}^{\xi_{\alpha}+z/\sqrt{n}}\left\{f_{t-1}(x)-f_{t-1}(\xi_{\alpha})\right\}dx\right|\leq K_{t-1} \frac{z^2}{\sqrt{n}},$$
by {\bf A1}.
Now, note that we have $E\frac{1}{\sqrt{n}}\sum_{t=1}^n\overline{\nu}_{t,n}=0$ and \begin{eqnarray*}\mbox{Var}\left(\frac{1}{\sqrt{n}}\sum_{t=1}^n\overline{\nu}_{t,n}\right)=E\overline{\nu}^2_{1,n}=
\int_{\xi_{\alpha}}^{\xi_{\alpha}+z/\sqrt{n}}E\left\{1-\frac{z}{\sqrt{n}}f_{0}(\xi_{\alpha})-\frac{k_{t,n}}{\sqrt{n}}\right\}^2f_{0}(x)dx\to 0,\end{eqnarray*}
using again {\bf A1}.
Moreover, almost surely
$$\frac{1}{\sqrt{n}}\sum_{t=1}^nE(\nu_{t,n}\mid {\cal F}_{t-1})=\frac{1}{\sqrt{n}}\sum_{t=1}^n\int_{\xi_{\alpha}}^{\xi_{\alpha}+z/\sqrt{n}}f_{t-1}(x)dx\to zEf_{t-1}(\xi_{\alpha}).$$
We thus have shown that $\sum_{t=1}^n\mathbf{1}^*_{\{u_t-\xi_\alpha\in (0,z/\sqrt{n})\}}=O_P(\sqrt{n})$.
Thus,  i) for $i=2$ is established.
By the same arguments and \eqref{diffhatet}, it can be  shown that i) for $i=3$ holds.

Turning to ii),
we have
\begin{eqnarray*}E\left(
u_t\mathbf{1}_{\{u_t-\xi_{\alpha}\in
(0,z/\sqrt{n})\}}^{*}D'_t\mid {\cal F}_{t-1}\right)&=&\int_{\xi_{\alpha}}^{\xi_{\alpha}+z/\sqrt{n}}xf_{t-1}(x)dx \bD_t'\\
&=&\int_{\xi_{\alpha}}^{\xi_{\alpha}+z/\sqrt{n}}xf_{t-1}(\xi_{\alpha})dx \bD_t'+
\int_{\xi_{\alpha}}^{\xi_{\alpha}+z/\sqrt{n}}x\{f_{t-1}(x)-f_{t-1}(\xi_{\alpha})\}dx \bD_t'\\
&=&
\xi_{\alpha}f_{t-1}(\xi_{\alpha})\frac{z}{\sqrt{n}}\bD_t'+
\frac{k^*_{t,n}}{\sqrt{n}}\bD_t'
\end{eqnarray*}
with
\begin{eqnarray*}\left|k^*_{t,n}\right|&=&\sqrt{n}\left| f_{t-1}(\xi_{\alpha})\frac{z^2}{2n}+\int_{\xi_{\alpha}}^{\xi_{\alpha}+z/\sqrt{n}}x\left\{f_{t-1}(x)-f_{t-1}(\xi_{\alpha})\right\}dx\right|\\&\leq & \frac{z^2}{\sqrt{n}}\left\{2K_{t-1}\xi_{\alpha}+\frac{f_{t-1}(\xi_{\alpha})}2+\frac{|z|}{\sqrt{n}}K_{t-1}\right\}.
\end{eqnarray*}

Denoting by $d_t$ a generic element of $\bD_t$, we also have
\begin{eqnarray}
\nonumber
&&EE\left[\left\{u_t\mathbf{1}_{\{u_t-\xi_{\alpha}\in
(0,z/\sqrt{n})\}}^{*}d_t-E\left(
u_t\mathbf{1}_{\{u_t-\xi_{\alpha}\in
(0,z/\sqrt{n})\}}^{*}d_t\mid {\cal F}_{t-1}\right)\right\}^2\mid {\cal F}_{t-1}\right]\\
&=&\int_{\xi_{\alpha}}^{\xi_{\alpha}+z/\sqrt{n}}E\left(x-\xi_{\alpha}f_{t-1}(\xi_{\alpha})\frac{z}{\sqrt{n}}+
\frac{k^*_{t,n}}{\sqrt{n}}\right)^2d_t^2f_{t-1}(x)dx =o(1),
\label{rousse}
\end{eqnarray}
as $n\to\infty$. To show that the expectation inside the latter integral is finite, we used in particular the fact that $$E\sup_{x\in[\xi_{\alpha},\xi_{\alpha}+z/\sqrt{n}]}d_t^2f^2_{t-1}(\xi_{\alpha})f_{t-1}(x)\leq \sqrt{Ed_t^8E\sup_{\xi\in V(\xi_{\alpha})}f^4_{t-1}(\xi)}<\infty$$
for sufficiently large $n$ under {\bf A1} and {\bf A7}. 
Hence, ii) is established.

To prove iii), write $J_n(z)=\sum_{t=1}^nJ_{n,t}$. Write $r_{n,t}=r_{n,t}(\hat{\btheta}_n)$, $R_{n,t}=R_{n,t}(\hat{\btheta}_n)$, $J_{n,t}=J_{n,t}(\hat{\btheta}_n)$ and $J_n(z)=J_n(z,\hat{\btheta}_n)$.
Let $(\btheta_n)$ be a deterministic sequence such that $\sqrt{n}(\btheta_n-\btheta_0)=O(1)$. By the
change of variable $u=u_tv$, we have
\begin{eqnarray*}
E \left(J_{n,t}(\btheta_n)\mid {\cal F}_{t-1}\right)
&=&
\int_0^{\bD_t'(\btheta_n-\btheta_0)+o_P(n^{-1/2})}E\left(u_t\mathbf{1}_{\{u_t\in
(\xi_{\alpha}+z/\sqrt{n},
(\xi_{\alpha}+z/\sqrt{n})(1-v)^{-1})\}}^{*}\mid {\cal F}_{t-1}\right)dv\\ &=&
\frac{\xi_{\alpha}^2}2f_{t-1}(\xi_{\alpha})({\btheta}_n-\btheta_0)'\bD_t\bD_t'({\btheta}_n-\btheta_0)+o_P(n^{-1}).
\end{eqnarray*}
By the arguments used to show \eqref{rousse}, we can show that
$$E\left[J_{n,t}(\btheta_n)- E\left\{J_{n,t}(\btheta_n)\mid {\cal F}_{t-1}\right\}\right]^2=o(n^{-1}).$$
We thus have
$$J_n(z,{\btheta}_n)=\frac{\xi_{\alpha}^2}2\sqrt{n}({\btheta}_n-\btheta_0)'E\{f_{0}(\xi_{\alpha})D_1D_1'\}\sqrt{n}({\btheta}_n-\btheta_0)+o(1),\quad a.s.$$
It follows that $J_n(z,{\btheta}_n)$ does not depend of $z$ asymptotically.
Since this is true for any sequence such that $\sqrt{n}(\btheta_n-\btheta_0)=O(1)$, this is also true almost surely for $J_n(z)$ and iii) is established.

By the previously used arguments, it can be shown that iv) holds which completes the proof of \eqref{yes}. 
By Lemma 2.2
in Davis et al. (1992) and convexity arguments, we can conclude that
\begin{eqnarray*} \sqrt{n}(\xi_{\alpha}-\xi_{n,
\alpha})&=&\frac{\xi_{\alpha}}{Ef_0(\xi_{\alpha})}\bPsi_{\alpha}'\sqrt{n}(\hat{\btheta}_n-\btheta_0)
+\frac1{Ef_0(\xi_{\alpha})}\frac{1}{\sqrt{n}}\sum_{t=1}^n
(\mathbf{1}_{\{u_t<\xi_{\alpha}\}}-\alpha)+o_P(1).
\end{eqnarray*}
We have the following Taylor expansion
\begin{eqnarray*} \sqrt{n}(\hat{\btheta}_n-\btheta_0)&=&
\frac{-\bJ^{-1}}{2\sqrt{n}}\sum_{t=1}^n
(1-u_t^2)\bD_t
+o_P(1).
\end{eqnarray*}
By the CLT for martingale differences we get the announced result, noting that
\begin{eqnarray*} \mbox{Cov}_{as}\left(\sqrt{n}(\hat{\btheta}_n-\btheta_0),
\frac{1}{\sqrt{n}}\sum_{t=1}^n
(\mathbf{1}_{\{u_t<\xi_{\alpha}\}}-\alpha)\right)
&=& \frac12\bJ^{-1}\bS_{\alpha}^{12},
\end{eqnarray*}
which entails
\begin{eqnarray*} &&\mbox{Var}_{as}\{\sqrt{n}(\xi_{n,
\alpha}-\xi_{\alpha})\}=\zeta_{\alpha},\qquad \mbox{Cov}_{as}\left(\sqrt{n}(\hat{\btheta}_n-\btheta_0),
\sqrt{n}(\xi_{\alpha}-\xi_{n,
\alpha})\right)=
\bLambda_{\alpha}.
\end{eqnarray*}
\zak

\section{Designs for the cDCC-GARCH model}\label{appendDCC}

The cDCC-GARCH(1,1) model is defined by
$\bSigma_t=\bD_t\bR_t^{1/2}$ where
the diagonal matrix $\bD_t=\mbox{diag}(\sigma_{1t},\dots,\sigma_{mt})$ is assumed to satisfy the GARCH(1,1) equation
\begin{equation}
\label{CCC}
\bh_t=\bomega_0+\bA_0
\underline{\by}_{t-1}  +\bB_0
\bh_{t-1}
\end{equation}
where $\bh_t= \left(\sigma_{1t}^2,\cdots,
\sigma_{mt}^2 \right)'$, $\underline{\by}_t= \left(\epsilon_{1t}^2,\cdots,
\epsilon_{mt}^2 \right)'$,
 $\bA_0$ and $\bB_0$ are  $m\times m$ matrices  with
positive coefficients, $\bomega_0$ is a vector of  strictly positive coefficients, and $\bB_0$ is assumed to be diagonal,
$$\bR_t=\bQ_t^{*-1/2}\bQ_t\bQ_t^{*-1/2},\quad
\bQ_t=(1-\alpha_0-\beta_0)\bS_0+\alpha_0 \bQ_{t-1}^{*1/2}\beeta_{t-1}^*\beeta_{t-1}^{*'}\bQ_{t-1}^{*1/2}+\beta_0\bQ_{t-1},$$
where $\alpha_0, \beta_0 \geq0, \alpha_0+\beta_0<1$, $\bS_0$ is a correlation matrix,
$\bQ_t^*$ is the diagonal matrix with the same diagonal elements as $\bQ_t$, and $\beeta_{t}^*=\bD_t^{-1}\by_t$.
The unknown parameter $\btheta_0$ includes the volatility parameters $\bomega_0$, $\bA_0$ and
$\mbox{diag}(\bB_0)$, and the conditional correlation parameters  $\alpha_0$, $\beta_0$ and the sub-diagonal elements of $\bS_0$.

\renewcommand{\baselinestretch}{1}\small\normalsize
\begin{center}
\begin{table}[!b]
\caption{\label{design}{\footnotesize
Design of Monte Carlo experiments.}}
\resizebox{1\textwidth}{!} {
\begin{tabular}{lcc ccc cc}
\hline\hline
 &$\bomega_0'$    & $(\mbox{vec}\bA_0)'$   & $\mbox{diag}\bB_0$   & $\bS_0(1,2)$   & $\alpha$  & $\beta$ &$P_{\beeta}$\\
A &$(10^{-6},\, 4\times 10^{-6})$    & $(0.01,\, 0.01,\, 0.01,\, 0.07)$   & $(0,\, 0.92)$   & $0.7$   & $0.04$  & $0.95$ &${\cal N}(0,\bI_2)$\\
B &$(10^{-6},\, 4\times 10^{-6})$    & $(0.01,\, 0.01,\, 0.01,\, 0.07)$   & $(0,\, 0.92)$   & $0.7$   & $0.04$  & $0.95$ &${\cal S}t_7$\\
C &$(10^{-6},\, 4\times 10^{-6})$    & $(0.01,\, 0.01,\, 0.01,\, 0.07)$   & $(0,\, 0.92)$   & $0$   & $0$  & $0$ &${\cal N}(0,\bI_2)$\\
D &$(10^{-6},\, 4\times 10^{-6})$    & $(0.01,\, 0.01,\, 0.01,\, 0.07)$   & $(0,\, 0.92)$   & $0$   & $0$  & $0$ &${\cal S}t_7$\\
E &$(10^{-5},\,  10^{-5})$    & $(0.07,\, 0.00,\, 0.00,\, 0.07)$   & $(0.92,\, 0.92)$   & $0.7$   & $0.04$  & $0.95$ &${\cal N}(0,\bI_2)$\\
F &$(10^{-5},\,  10^{-5})$    & $(0.07,\, 0.00,\, 0.00,\, 0.07)$   & $(0.92,\, 0.92)$   & $0.7$   & $0.04$  & $0.95$ &${\cal S}t_7$\\
G &$(10^{-5},\,  10^{-5})$    & $(0.07,\, 0.00,\, 0.00,\, 0.07)$   & $(0.92,\, 0.92)$   & $0$   & $0$  & $0$ &${\cal N}(0,\bI_2)$\\
H &$(10^{-5},\, 10^{-5})$    & $(0.07,\, 0.00,\, 0.00,\, 0.07)$   & $(0.92,\, 0.92)$   & $0$   & $0$  & $0$ &${\cal S}t_7$\\
\hline
\multicolumn{8}{l}{Designs A$^*$-H$^*$ are the same as Designs A-H, except that $P_{\beeta}$ follows an  AEPD. }
\end{tabular}}
\end{table}
\end{center}

\renewcommand{\baselinestretch}{1.4}\small\normalsize
The parameters used in the Monte-Carlo experiments of Section \ref{secestmul} are displayed in Table \ref{design}.
In Designs A-D the first return is less volatile and less conditionally heteroscedastic than the second return, whereas the two returns have the same dynamic in Designs E-H. Two sets of designs are also distinguished by strong dynamic correlations ($\alpha_0+\beta_0=0.99$) with a strong correlation between the returns ($\bS_0(1,2)=0.7$) or constant conditional correlations with null cross-correlation ($\alpha_0=\beta_0=0$ and $\bS_0(1,2)=0$). Finally, the designs are distinguished by the distribution of the innovations, which can be the standard normal or the Student distribution with 7 degrees of freedom ${\cal S}t_7$ (standardized to obtain unit variance).
For generating non spherical distributions, we simulated vectors $\beeta_t$ with independent components, distributed according to the  Asymmetric Exponential Power Distribution (AEPD) introduced by Zhu and Zinde-Walsh  (2009). This class of distributions allows for skewness with different decay rates of density in the left and right tails.   This led to the new Designs A$^*$-H$^*$, in which the  ${\cal N}(0,\bI_2)$ is replaced by the AEPD with parameters $\alpha=0.4$, $p_1=1.182$ and $p_2=1.802$ (which are the values estimated by Zhu and Zinde-Walsh  on the S\&P500),
and the Student distribution ${\cal S}t_7$ is replaced by the AEPD with parameters $\alpha=0.5$, $p_1=1$ and $p_2=2$ (which gives a strongly asymmetric density).
The AEPD densities have also been standardized to obtain zero mean and unit variance.

\bigskip
\noindent{\large {\bf Acknowledgements}

We thank S. Darolles and C. Hurlin for helpful discussions.
We are  grateful to the Agence Nationale de la
Recherche (ANR), which supported this work via the Project
MultiRisk  (ANR CE26 2016 - CR), and the Labex ECODEC.
The second author gratefully thanks the IDR ACP "Régulation et risques systémiques" for financial support.

\vspace{1em}
\newpage
\begin{center}
{\Large\bf References}
\end{center}
\begin{description}
\item[Aielli, G.P.] (2013)
Dynamic conditional correlation: on properties and estimation.
{\it Journal of Business \& Economic Statistics} 31, 282--299.
\item[Andrews, D.W.K.] (1991) Heteroskedasticity and autocorrelation consistent covariance matrix estimation. {\it Econometrica} 59,
 817--858.
\item[Barone-Adesi, G., Giannopoulos, K., and L. Vosper] (1999)
VaR without correlations for nonlinear portfolios.
{\it Journal of Futures Markets} 19, 583--602.
\item[Bauwens, L.,  Hafner, C.M. and S. Laurent] (2012)
{\it Handbook of Volatility Models and Their Applications.}
Wiley.
\item[Bauwens, L.,  Laurent, S. and J.V.K. Rombouts] (2006)
 Multivariate GARCH models: a survey. {\it Journal of Applied
 Econometrics} 21, 79--109.
\item[Beutner, B., Heinemann, A. and S. Smeekes] (2019)
A justification of conditional confidence intervals. Discussion paper. arXiv:1710.00643v2.
\item[Beutner, B., Heinemann, A., and S. Smeekes] (2018)
 A residual bootstrap for conditional value-at-risk. Discussion paper. arXiv:1808.09125.
\item[Billingsley, P.](1999)
{\it Convergence of Probability Measures}. 2nd edition, Wiley.
\item[Boyd, S., and L. Vandenberghe] (2018) {\it Introduction to Applied Linear Algebra: Vectors, Matrices, and Least Squares.} Cambridge University Press.
\item[Boudt, K., Laurent, S., Quaedvlieg, R. and O. Sauri] (2017)
Positive semidefinite integrated covariance estimation, factorizations and asynchronicity.
{\it Journal of Econometrics} 196, 347--367.
\item[Christoffersen, P.F.]  (2003)
{\it Elements of financial risk management.} Academic Press,
London.
\item[Christoffersen, P.F. and S. Gonçalves]  (2005)
Estimation risk in financial risk management.
{\it Journal of Risk}  7, 1--28.
\item[Davis, R.A., Knight, K.  and J. Liu] (1992)
M-estimation for autoregressions with infinite variance. {\it
Stochastic Processes and their Applications} 40, 145--180.
\item[Diebold, F.X. and R.S. Mariano] (1995) Comparing predictive accuracy. {\it Journal of Business  \& Economic Statistics} 13, 253--263.
\item[Engle, R.F., Ledoit, O. and  M. Wolf] (2017)
Large dynamic covariance matrices.
{\it Journal of Business \& Economic Statistics} DOI: 10.1080/07350015.2017.1345683
\item[Escanciano, J.C., and J. Olmo] (2010) Backtesting parametric VaR with estimation
 risk. {\it Journal of Business \& Economic Statistics} 28, 36--51.
\item[Escanciano, J.C. and J. Olmo] (2011)
Robust backtesting tests for value-at-risk models.
{\it  Journal of Financial Econometrics} 9, 132--161.
\item[Farkas, W., Fringuellotti, F. and R. Tunaru] (2016)
Regulatory capital requirements: saving too
much for rainy days? Unpublished document, University of Zürich.
\item[Francq, C., and J.M. Zakoïan] (2005) A central limit theorem for mixing triangular arrays
of variables whose dependence is allowed to grow with the sample size.
 {\it Econometric Theory} 21, 1165--1171.
\item[Francq, C.,  and J.M. Zakoïan] (2010)
{\it GARCH models: structure, statistical inference and
financial applications.} Chichester: John Wiley.
\item[Francq, C. and J.M. Zakoïan] (2013)
 Optimal predictions of powers of conditionally
heteroskedastic processes. {\it Journal of the Royal Statistical Society - Series B} 75, 345--367.
\item[Francq, C. and J.M. Zakoïan] (2015)
Risk-parameter estimation in volatility models. {\it Journal of Econometrics} 184, 158--173.
\item[Francq, C. and J.M. Zakoïan] (2016)
Estimating multivariate GARCH models equation by equation.
{\it Journal of the Royal Statistical Society: Series B (Statistical Methodology)} 78, 613--635.
\item[Francq, C. and J.M. Zakoïan] (2018)
Estimation risk for the VaR of portfolios driven by semi-parametric multivariate models.
 {\it Journal of Econometrics} 205, 381--401.
\item[Giacomini, R. and I. Komunjer] (2005) Evaluation and combination of conditional quantile
forecasts. {\it Journal of Business \& Economic Statistics} 23, 416--431.
\item[Gneiting, T.] (2011)
Making and evaluating point forecasts. {\it Journal of the American Statistical Association} 106,
746--762.
\item[Gong, Y., Li, Z. and L. Peng] (2010)
Empirical likelihood intervals for conditional value-at-Risk in ARCH/GARCH models.
{\it Journal of Time Series Analysis} 31, 65--75.
\item[Gouriéroux, C. and J.M. Zakoïan] (2013)
Estimation adjusted VaR.  {\it Econometric Theory} 29, 735--770.
\item[Hansen, B.E.] (2004)
Nonparametric conditional density estimation.
Discussion paper, University of Wisconsin.
\item[Holton, G.A.]  (2014) {\it  Value-at-Risk: Theory and Practice}. Second Edition, Academic press.
\item[Hurlin, C., Laurent, S., Quaedvlieg, R. and S. Smeekes](2017)
Risk measure inference. {\it Journal of Business \& Economic Statistics} 35, 499--512.
\item[Knight, K.] (1998) Limiting distributions for $L_1$ regression estimators under general conditions. {\it The Annals of Statistics} 26, 755--770.
\item[Koenker, R. and Z. Xiao] (2006) Quantile autoregression.
{\it Journal of the American Statistical Association} 101,
980--990.
\item[Laurent, S., Lecourt, C. and F.C. Palm]  (2016)
Testing for jumps in conditionally Gaussian ARMA-GARCH models, a robust approach. {\it Computational Statistics \& Data Analysis} 100, 383--400.
\item[Laurent, J.P., and H. Omidi Firouzi] (2017) Market risk and volatility weighted historical simulation after Basel III. {\it Preprint} available at \url{https://www.msci.com/documents/10199/5915b101-4206-4ba0-aee2-3449d5c7e95a}
\item[Mancini, L. and F. Trojani] (2011)
Robust Value-at-Risk prediction. {\it
Journal of Financial Econometrics} 9, 281--313.
\item[Newey, W.K and K.D. West] (1987) A simple, positive semi-definite, heteroskedasticity and autocorrelation consistent
covariance matrix. {\it Econometrica} 55,  703--708.
\item[Nieto, M. R. and E. Ruiz] (2016)
Frontiers in VaR forecasting and backtesting.
{\it International Journal of Forecasting} 32, 475--501.
\item[Pérignon, C., Deng, Z.Y., and Z.J. Wang] (2008) Do banks overstate their Value-at-Risk? {\it Journal of Banking \& Finance} 32, 783--794.
\item[Rombouts, J.V.K. and M. Verbeek] (2009)
Evaluating portfolio Value-at-Risk using semi-parametric GARCH models. {\it Quantitative Finance} 9, 737--745.
\item[Santos, A.A.P., Nogales F.J. and E. Ruiz] (2013)
Comparing univariate and multivariate models to forecast portfolio Value-at-Risk.
{\it Journal of Financial Econometrics} 11, 400--441.
\item[Spierdijk, L.] (2016)
Confidence intervals for ARMA-GARCH Value-at-Risk:
the case of heavy tails and skewness.
{\it Computational Statistics \&        Data Analysis} 100, 545--559.
\item[Zhu, D. and V. Zinde-Walsh] (2009) Properties and estimation of asymmetric exponential power distribution.
{\it Journal of Econometrics} 148, 86--99.
\end{description}

\end{document}